\documentstyle[prd,aps,preprint,psfig]{revtex}

\newcommand{\hefour}{{\rm ^4He}}

\newcommand{\liseven}{{\rm ^7Li}}

\newcommand{\kev}{ {\rm keV} }
\newcommand{\mev}{ {\rm MeV} }
\newcommand{\gev}{ {\rm GeV} }
\newcommand{\tev}{ {\rm TeV} }
\newcommand{\mb}{ {\rm mb} }
\newcommand{\order}{{\cal O}}

\newcommand{\anti}[1]{{\overline{#1}}}
\newcommand{\tenten}{ {\cdot\cdot\cdot} }

\begin{document}
\tighten

\preprint{
\noindent
\begin{minipage}[t]{3in}
\begin{flushright}
YITP-01-23\\
astro-ph/0103411 \\
May 2001\\
\end{flushright}
\end{minipage}
}

\title{Primordial nucleosynthesis and hadronic decay of a massive
particle with a relatively short lifetime}
\author{Kazunori Kohri}
\address{Yukawa Institute for Theoretical Physics, Kyoto University,
Kyoto, 606-8502, Japan}

\maketitle
\begin{abstract}
    In this paper we consider the effects on big bang nucleosynthesis
    (BBN) of the hadronic decay of a long-lived massive particle. If
    high-energy hadrons are emitted near the BBN epoch ($t \sim
    10^{-2}$ -- $10^2 \sec$), they extraordinarily inter-convert the
    background nucleons each other even after the freeze-out time of
    the neutron to proton ratio. Then, produced light element
    abundances are changed, and that may result in a significant
    discrepancy between standard BBN and observations. Especially on
    the theoretical side, now we can obtain a lot of experimental data
    of hadrons and simulate the hadronic decay process executing the
    numerical code of the hadron fragmentation even in the high energy
    region where we have no experimental data. Using the computed
    light element abundances in the hadron-injection scenario, we
    derive a constraint on properties of such a particle by comparing
    our theoretical results with observations.
\end{abstract}



\section{Introduction}

Big bang nucleosynthesis (BBN) is one of the most important tools to
probe the early universe because it is sensitive to the condition of
the universe from $10^{-2}$ sec to $10^{4}$ sec. Therefore, from the
theoretical predictions we can indirectly check the history of the
universe in such an early epoch and impose constraints on hypothetical
particles by observational light element abundances.

Now we have a lot of models of modern particle physics beyond the
standard model, e.g., supergravity or superstring theory, which
predict unstable massive particles with masses of $\order$(100) GeV --
$\order$(10) TeV, such as the gravitino, Polonyi field, and moduli.
They have long lifetimes because their interactions are suppressed by
inverse powers of the gravitational scale ($\propto 1/M_{\rm
pl}$). These exotic particles may necessarily decay at about the BBN
epoch ($T \lesssim \order(1)$ MeV) if they have already
existed in earlier stages. If the massive particles radiatively decay,
the emitted high energy photons induce the electromagnetic cascade
process. If the decay occurs after BBN starts $t \gtrsim 10^4 \sec$,
the light elements would be destroyed by the cascade photons and their
abundances would be changed significantly.  Comparing the theoretical
prediction of light element abundances with the observations, we can
impose constrains on the energy density, the mass, and the lifetime of
the parent massive particle~\cite{Ellis:1992nb,Kawasaki:1995af,Holtmann:1999gd}. This subject was
also studied in more details in the recent paper~\cite{Kawasaki:2001qr}.

On the other hand, if the massive particles decay into quarks or
gluons, near the BBN epoch $10^{-2} \lesssim t \lesssim 10^2 \sec$. it
is expected that the other important effects are induced. If once
the high energy quarks or gluons are emitted, they quickly fragment
into a lot of hadrons. Then, such high energy hadrons are injected
into the electromagnetic thermal bath which is constituted by photons,
electrons, and nucleons (protons and nucleons) at that time. At first,
the high energy hadrons scatter off the background photons and
electrons because they are more abundant than the background nucleons.
Then, almost all kinetic energy of the hadrons are transfered into the
thermal bath through the electromagnetic interaction. As a result,
they are completely stopped and reach to the kinetic
equilibrium. After that time, they scatter off the background $p$ or
$n$ through the strong interaction, and they inter-convert the
background $p$ and $n$ each other even after the usual freeze-out time
of the neutron to proton ratio $n/p$ of the weak interaction. The
effect extraordinarily tends to increase $n/p$. Therefore, the
produced $\hefour$ would be increased in the hadron injection scenario
compared to standard big-bang nucleosynthesis (SBBN).

The pioneering investigation of this subject was done by Reno and
Seckel (1988)~\cite{Reno:1988qw}, and their treatments have been also applied
to the other subjects~\cite{Kawasaki:2000en,Kohri:2000ex}. After their work was published,
the experiments of the high energy physics have been widely developed.
Now we can obtain a lot of experimental informations of the hadron
fragmentation in the high energy region and also simulate the process
even in the higher energies where we have no experimental data by
executing the numerical code of the hadron fragmentation, e.g. JETSET
7.4 Monte Carlo event generator~\cite{Sjostrand:1994yb}. In addition, we have
more experimental data of the hadron-nucleon cross
sections. Concerning BBN computations, it is recently needed that we
perform a Monte Carlo simulation which includes the experimental
errors of the reactions, and then we estimate the confidence levels
(C.L.) by performing the Maximum Likelihood analysis and the $\chi^2$
fitting including both the theoretical and the observational
errors. Performing the above procedures, we can compare each model in
the various parameter sets. With these new developments in the theory,
we set bounds to
the hadronic decay of long-lived particles.~\footnote{
For relatively longer lifetimes, there exists an another interesting
process that the emitted high energy nucleons destroy the light
elements which have already been produced~\cite{DEHS}.
}

This paper is organized as follows. In Sec.~II we briefly review the
current status of the observations and SBBN. In Sec.~III we introduce
the formulations and computations in the hadron injection scenario. In
Sec.~IV we compare the theoretical predictions with the
observations. Section V is devoted to the summary and conclusions.

\section{current status of observational light element abundances and SBBN}
\label{sec:obs_status}

\subsection{Current status of observations}

In this section, we briefly summarize the current status of the
observational light element abundances.  The primordial D/H is
measured in the high redshift QSO absorption systems.  Recently a new
deuterium data was obtained from observation of QSO HS 0105+1619 at z
= 2.536~\cite{ap0011179}.  The reported value of the deuterium
abundance was relatively low, (D/H)$^{obs} = ( 2.54 \pm 0.23 ) \times
10^{-5}$.  Combined with the previous ``low D'' data~\cite{BurTyt},
the authors reported that the primordial abundance is
\begin{equation}
      \label{lowd}
     {\rm low D} : \left( {\rm D/H} \right)^{obs}
     = (3.0 \pm 0.4) \times 10^{-5}.
\end{equation}
We call this value ``low D.'' On the other hand, Webb et al. obtained
high deuterium abundance in relatively low redshift absorption systems
at z = 0.701 towards QSO PG1718+4807~\cite{webb},
\begin{equation}
      \label{eq:highd} 
      {\rm high D} : \left( {\rm D/H} \right)^{obs} = (2.0 \pm 0.5)
      \times 10^{-4}.
\end{equation}
In these days, Kirkman et al.~\cite{kirkman01} also observed the
clouds independently and obtained new spectra using HST. They claimed
that the absorption was not deuterium although there were still some
uncertainties. Here we think that it is premature to decide which
component is correctly primordial and the possibility of ``high D''
have not been excluded yet.  Therefore, we also consider the
possibility of ``high D'' and include it in our analysis.

The primordial value of $^4$He is inferred from the recombination
lines from the low metallicity extragalactic HII regions.  The
primordial value of $^4$He mass fraction $Y$ is obtained to regress to
the zero metallicity O/H $\rightarrow 0$ for the observational data
because it is produced with oxygen in stars.  In these days Fields and
Olive reanalyzed the data including the HeI absorption
effect~\cite{FieOLi}. Then they obtain the observational $Y$,
\begin{equation}
      \label{FieOLi}
       Y^{obs} = 0.238 \pm (0.002)_{stat} \pm (0.005)_{syst},
\end{equation}
where the first error is the statistical uncertainty and the second
error is the systematic one. We adopt this value as the
observational value of $Y$.

It is widely believed that the primordial abundance of $^7$Li/H is
observed in the Pop II old halo stars whose temperature is high
$T_{\rm eff} \gtrsim 6000$ K and metallicity is low [Fe/H] $\lesssim$ -
1.5.  They have the ``plateau'' structure of $^7$Li/H as a function of
the metallicity. We adopt the recent measurements by Bonifacio and
Molaro~\cite{BonMol}
\begin{equation}
      \label{eq:li7}
      {\rm log_{10}}\left[ \left(\liseven/{\rm H}\right)^{obs} \right]
      =-9.76 \pm (0.012)_{stat} \pm (0.05)_{syst} \pm (0.3)_{add}.
\end{equation}
Here we have added the additional uncertainty for fear that the $\liseven$
in halo stars might have been supplemented (by production in
cosmic-ray interactions) or depleted (in
stars)~\cite{factor-of-two}.~\footnote{
These days, however, it was claimed that there is a significant Li-Fe
trend in the low metallicity region~\cite{RNB}. In addition, Ryan et
al.~\cite{RBOFN} assumed that this trend is due to the cosmic ray
interactions, and they inferred the primordial value is $^7$Li/H $=
(1.23^{+0.68}_{-0.32}) \times 10^{-10}$.  Because we can not make a
judgment on the above discussions, for the moment we adopt the value
in Eq.~(\ref{eq:li7}) with large uncertainties in this paper.
}

\subsection{Current status of SBBN}
\label{subsec:SBBN}

Here we show the current status of standard big-bang nucleosynthesis
(SBBN). Within recent years, there was a great progress in the
experiments of the low energy cross sections for 86 charged-particle
reactions by the NACRE collaboration~\cite{NACRE}. In the compilation,
22 reactions are relevant to the primordial nucleosynthesis, and the
old data were revised. In particular, 7 reactions of them are
important for the most elementary processes up to mass-7 elements. On
the other hand, recently Cyburt, Fields and Olive reanalyzed the NACRE
data and properly derived the $1\sigma$ uncertainty as a statistical
meaning and the renormalization of the center value for each
reaction~\cite{CFO}. In addition, they also reanalyzed the four
remaining reactions, using the existing data~\cite{SKM,kawano,Brune}
and the theoretical prediction (for one reaction)~\cite{Hale}. Their
efforts are quite useful for the study of the Monte Carlo simulation
in BBN, and it was shown that their treatment is consistent with the
other earlier studies adopting the results of NACRE~\cite{NB,VCC}.

Carrying the Monte Carlo simulation into execution, we adopt the
theoretical errors and the center values for 11 elementary nuclear
reactions in Ref.~\cite{CFO}. For the error and the center value of
neutron lifetime, we adopt the compilation of Particle Date
Group~\cite{PDG}, see Eq.~(\ref{eq:tau_n}). To systematically take
account of the uncertainties, we perform the Maximum Likelihood
analysis~\cite{Holtmann:1999gd} including both the observational and theoretical
errors which are obtained in Monte Carlo simulation. Here we assume
that the theoretical predictions of (D/H)$^{th}$, $Y^{th}$, ${\rm
log_{10}}[(\liseven/{\rm H})^{th}]$ obey the Gaussian probability
distribution functions (p.d.f.'s) with the widths given by the
1$\sigma$ errors. Concerning the observational values, (D/H)$^{obs}$,
$Y^{obs}$, and ${\rm log_{10}}[(\liseven/{\rm H})^{obs}]$ are also
assumed to obey the Gaussian p.d.f.'s.

\begin{figure}
      \begin{center}
          \centerline{\psfig{figure=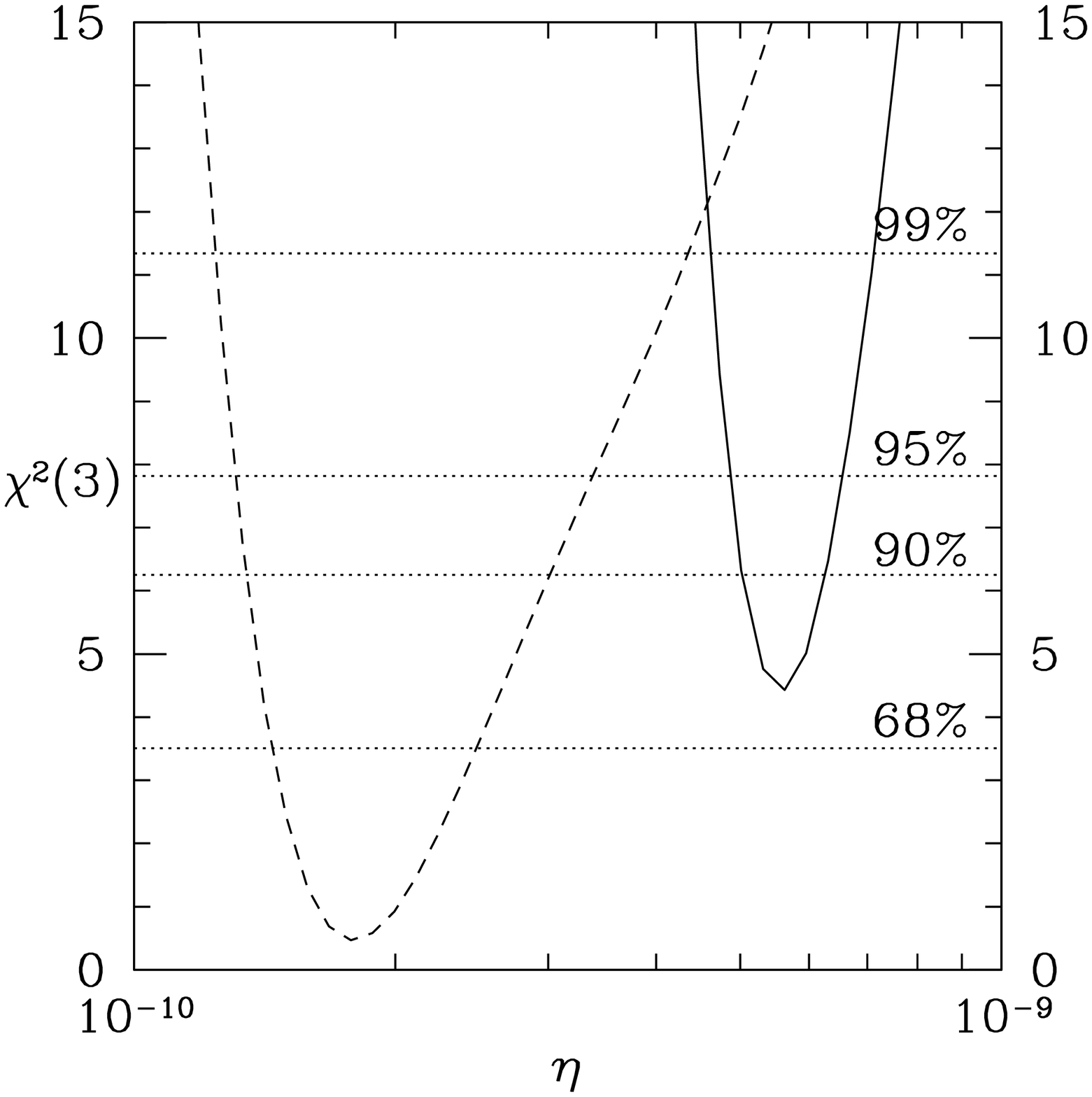,width=17cm}}
          \caption{
          Plot of $\chi^2$ as a function of baryon to photon ratio
          ($\eta = n_B/n_{\gamma}$). The solid line (dashed line)
          represents the case of low D (high D).
          }
          \label{fig:chi_2001_cfo}
      \end{center}
\end{figure}

In Fig.~\ref{fig:chi_2001_cfo} we plot $\chi^2$ as a function of
baryon to photon ratio, $\eta = n_B/n_{\gamma}$, where $n_B$ is the
baryon number, and $n_{\gamma}$ is the photon number. The solid line
(dashed line) represents the case of low D (high D). From this figure,
we find that SBBN agrees with the observation of $\hefour$, D, and
$\liseven$ very well at more than 95 $\%$ C.L., and we obtain $\eta =
5.6^{+0.9}_{-0.8} \times 10^{-10}$ ($\eta = 1.8^{+1.6}_{-0.5} \times
10^{-10}$) for low D (high D) at 95 $\%$ C.L. Using the relation
$\Omega_B h^2 = 3.63 \times 10^7 \eta (T_0/2.725 K)$, we obtain 
\begin{eqnarray}
    \label{eq:omegab}
    \Omega_B h^2 = \left\{
        \begin{array}{ll}
            0.0203^{+0.0033}_{-0.0029}\qquad {\rm (for \ low D)},  \\
            \\
            0.0065^{+0.0058}_{-0.0018}\qquad {\rm (for \ high D)},
        \end{array}
      \right.
\end{eqnarray}
at 95 $\%$ C.L., where $\Omega_B$ is baryon density parameter, $h$ is
normalized Hubble parameter as $H_0$ = 100$h$ km/sec/Mpc, and $T_0$ is
the present temperature\cite{PDG}. Under these circumstances, we can
check the non-standard scenario comparing the predictions of the BBN
computations with observations.

\section{Hadronic decay and BBN}
\label{sec:hadron_BBN}

In this section, we discuss the hadron-injection effects on the
history of the universe near BBN epoch ($t = 10^{-2}$ -- $10^4$
sec). Here we consider the case that the unstable massive particle
``$X$'' has some decay modes into quarks and gluons, and as a result it
induces the late-time hadron injection.

\subsection{Time scale of the interactions}
\label{subsec:timescale}

If the quarks and gluons were emitted by the decay of the parent
particle $X$ whose mass is about $\order(100)$ GeV -- $\order(10)$
TeV, they immediately fragment into hadron jets and produce a lot of
mesons and baryons ($\pi^{\pm}, \pi^0,K^{\pm}, K^0_{L,S}, n, p,
\Lambda^0$, and so on).  Then, the typical energy of the produced
hadrons is about $\order(1)\gev$ -- $\order(100) \gev$, and they are
injected into the electromagnetic thermal bath which is constituted by
$\gamma, e^{\pm}$, and nucleons.

As we see later, if once such high energy hadrons are injected into
the thermal bath in the beginning of the BBN epoch (i.e., at the
temperature $T \gtrsim 0.09$ MeV), their almost all kinetic energy is
transfered into the thermal bath through the electromagnetic
interactions except for neutral kaons. Then, such hadrons scatter off
the background particles, and then they induce some non-standard
effects on BBN.  Especially, the emitted hadrons extraordinarily
inter-convert the ambient protons and neutrons each other through the
strong interaction even after the freeze-out time of the neutron to
proton ratio $n/p$.  For the relatively short lifetime ($\tau_{\phi}
\simeq 10^{-2} \sec$ -- $10^2 \sec$) in which we are interested, the
above effect induces the significant change in the produced light
elements. Concretely, protons which are more abundant than neutrons,
are changed into neutrons through the hadron-proton collisions and the
ratio $n/p$ increases extremely.  In this case, the late-time hadron
injection scenario tends to increase $\hefour$ because it is the most
sensitive to the freeze out value of $n/p$,

The emitted hadrons do not scatter off the background nucleons
directly. At first hadrons scatter off the background photons and
electrons because they are much more abundant than background nucleons
(about 10$^{10}$ times larger). As we see later, for $t \lesssim 200
\sec$, the emitted high energy hadrons are immediately thermalized
through the electromagnetic scattering and they reach to the kinetic
equilibrium before they interact with the ambient protons, neutrons
and light elements. Then we use the thermal-averaged cross section
$\langle\sigma v\rangle^{H_i}_{N \rightarrow N'}$ for the strong
interaction process $N + H_i \rightarrow N' + \cdot \cdot \cdot$
between hadron $H_i$ and the ambient nucleon $N$, where $N$ denotes
proton $p$ or neutron $n$. The strong interaction rate is estimated by
\begin{eqnarray}
     \label{eq:gamma^i_nn}
     \Gamma^{H_i}_{N\rightarrow N'} &=& n_N \langle\sigma
     v\rangle^{H_i}_{N \rightarrow N'} \nonumber \\ &\simeq& \left(2
       \times 10^{-8} \sec \right)^{-1} f_N
     \left(\frac{\eta}{10^{-9}}\right) \left(\frac{\langle\sigma
       v\rangle^{H_i}_{N \rightarrow N'}}{40 \mb} \right)
     \left(\frac{T}{1 \mev}\right)^3,
\end{eqnarray}
where $n_N$ is the number density of the nucleon species $N$, $\eta$
is the baryon to photon ratio ($=n_B/n_{\gamma}$), $n_B$ denotes the
baryon number density ($= n_p + n_n$), and $f_N$ is the nucleon
fraction ($ \equiv n_N/n_B$). Here, for the moment we adopt 40 mb as a
typical value of the cross section for the strong interaction. This
equation shows that every hadron whose lifetime is longer than ${\cal
O}(10^{-8})$ sec contributes to the inter-converting interaction
between neutron and proton at the beginning of BBN. Hereafter we will
consider only the long-lived mesons ( $\pi^{\pm}$, $K^{\pm}$, and
$K_L$) and
baryons ($p$, $\overline{p}$, $n$, and $\overline{n}$).~\footnote{
$\pi^0, K_S^0$, and $\Lambda^0$ have much shorter lifetimes and they
have completely finished to decay because their lifetimes are
$\tau_{\pi^0} = (8.4 \pm 0.6) \times 10^{-17} \sec$, $\tau_{K_S^0} =
0.89 \times 10^{-10}$, and $\tau_{\Lambda^0} = 2.63 \times 10^{-10}$
sec respectively. Therefore, they do not contribute to the interesting
process in this situation.
}
 Their lifetimes are given by~\cite{PDG}
\begin{eqnarray}
    \label{eq:lifetimes}
    \tau_{\pi^{\pm}} &=& (2.6033 \pm 0.0005) \times 10^{-8} \sec, \\
    \tau_{K^{\pm}} &=& (1.2386 \pm 0.0024) \times 10^{-8} \sec, \\
    \tau_{K^0_L} &=& (5.17\pm 0.04) \times 10^{-8} \sec, \\
    \label{eq:tau_n}
    \tau_n &=& 886.7 \pm 1.9 \sec, 
\end{eqnarray}
and proton is stable.

Here we define the stopping time $\tau^{H_i}_{\rm stop}$ of the high
energy particle $H_i$ in the thermal plasma as
\begin{eqnarray}
    \label{eq:tau-stop}
    \tau^{H_i}_{\rm stop} = \int^{E_{\rm th}}_{E_0} \left
    ( \frac{dE}{dt}\right)^{-1} dE,
\end{eqnarray}
where $E$ denotes the energy, $dE/dt$ denotes the energy loss rate in
the thermal plasma and it depends on the each scattering process of
particle $H_i$ off the background particles. $E_0$ is the initial
energy, and $E_{\rm th}$ is the threshold energy of the
process.~\footnote{
To roughly estimate the timescale until the particle is stopped, it
would be usually adequate that we take $E_{\rm th}$ to be equal to the
mass of the particle $H_i$ in the relativistic regime.
}
 To estimate whether particle $H_i$ is stopped or not in the thermal
plasma through the electromagnetic interaction until it scatter off
the background baryons ($n$, $p$, and produced light elements), we
computes the rate,
\begin{eqnarray}
    \label{eq:r_stop}
    R^{H_i}_{\rm stop} \equiv \Gamma^{H_i}_{N\rightarrow N'} \times
    \tau^{H_i}_{\rm stop},
\end{eqnarray}
as an indicator which roughly represents the number of the scattering
during the stopping time $\tau^{H_i}_{\rm stop}$.  If $R^{H_i}_{\rm
stop}$ is much less than unity, the emitted high energy hadron $H_i$
is completely stopped and can not reach the background baryons with
the high energy. On the other hand, if $R^{H_i}_{\rm stop}$ is greater
than unity, the high energy hadron can not be stopped through the
electromagnetic interaction and directly scatters off the background
baryons. In addition, it might destroy the light elements which have
already produced if the particle $X$ decays after the cosmic time is
$t \sim 200$ sec.

\subsection{Hadron stopping in the electromagnetic thermal plasma}
\label{subsec:stopping}

When the cosmic temperature $T$ is higher than the electron mass
$m_e$, there are sufficient electrons and positrons in the
universe. In this situation, it is expected that the emitted charged
particles $\pi^{\pm}, K^{\pm}$, and $p$ are quickly thermalized
through the electromagnetic interaction. In fact, the energy loss rate
of the charged particle through the Coulomb scattering is given by
\begin{eqnarray}
    \label{eq:dedt_ch_rel_high}
    \frac{dE}{dt} = - \frac{\pi}{3} \alpha T^2,
\end{eqnarray}
for $T \gtrsim m_e$ in the relativistic regime. $\alpha$ is fine
structure constant ($\simeq 1/137$).  Then, the stopping time of the
charged particle (``ch'') is estimated by
\begin{eqnarray}
    \label{eq:stop_ch_c_rel_high}
    \tau_{\rm stop}^{\rm ch} \simeq 1.18 \times 10^{-14} \sec
    \left(\frac{E}{\gev}\right)\left( \frac{T}{\mev}\right)^{-2},
\end{eqnarray}
for $T \gtrsim m_e$. Then, $R^{\rm ch}_{\rm stop}$ is much smaller
than unity and we can regard that charged hadrons are completely
stopped.

As for neutron, we can see that it is also completely stopped for $T
\gtrsim m_e$. Although neutron is neutral, of course, it can scatter
off the background electrons through the electromagnetic interaction
by the magnetic dipole moment. The energy loss rate through the
Coulomb scattering is given by
\begin{eqnarray}
    \label{eq:dedt_nc_rel_high}
    \frac{dE}{dt} = - \frac{15 m_n^3}{7 \pi^3 \alpha^2 g_n^2 T^4},
\end{eqnarray}
in the relativistic regime, where $g_n$ is neutron magnetic moment (=
$- 1.913$)~\cite{PDG}, and $m_n$ is neutron mass. The stopping time of
neutron is
\begin{eqnarray}
    \label{eq:stop_nc_c_rel_high}
    \tau_{\rm stop}^{\rm n} \simeq 2.34 \times 10^{-10}\sec
    \left(\frac{T}{\mev}\right).
\end{eqnarray}
Thus $R^{\rm n}_{\rm stop}$ of neutron is much smaller than unity, and
it does not scatter off the background baryons before it stops for $T
\gtrsim m_e$.~\footnote{
Although the above estimations have been discussed only in the
relativistic regime, the similar results are also obtained in the
non-relativistic regime~\cite{Reno:1988qw}.
}

On the other hand, if the temperature is much lower than electron mass
($T \lesssim m_e$), the situation is quite different because the
number density of electrons becomes little. In this case, the emitted
mesons completely decay and disappear in the universe before they
scatter off the background baryons because the lifetime is shorter
than the timescale of the strong interaction (see
Eq.~(\ref{eq:gamma^i_nn})).  Thus we should not treat the injection of
any mesons in such a late epoch. Because proton is stable, and neutron
has a long lifetime compared to the typical timescale of the strong
interaction in Eq.~(\ref{eq:gamma^i_nn}), we should worry about the
thermalization of the emitted high-energy nucleons till quite later.

For proton, the ionization loss is more effective to lose the
relativistic energy for $T \lesssim m_e$. The ionization-loss rate is
expressed by
\begin{eqnarray}
    \label{eq:dedt_ch_ion_rel}
    \frac{dE}{dt} = - \frac{Z^2 \alpha}{v} \omega_p^2
    \ln\left(\frac{\Lambda m_e \gamma v^2}{\omega_p}\right),
\end{eqnarray}
where $Z$ denotes the charge ($Z=1$ for proton), $v$ is the
velocity of the high energy proton, $\gamma$ is the Lorentz factor,
$\Lambda$ is $\order$(1) constant, and
$\omega_p$ denotes the plasma frequency,
\begin{eqnarray}
    \label{eq:omega_p}
    \omega_p^2 = \frac{4\pi \alpha n_e }{m_e},
\end{eqnarray}
where $n_e$ represents the electron number density. We evaluate the
stopping time of the proton to lose its relativistic energy,
\begin{eqnarray}
    \label{eq:stop_ch_ion_rel}
    \tau^p_{\rm stop} \simeq 1.2 \times 10^{-14} \sec x^{\frac12} e^x
    \left(\frac{E}{\gev}\right) \left(\frac{\eta_{10}}{5}\right)^{-1},
\end{eqnarray}
where $\eta_{10}$ is defined by $\eta = \eta_{10} \times 10^{-10}$,
and the dimensionless parameter $x = m_e/T$. If we demand $R^p_{\rm
stop} \lesssim 1$, we obtain $T \gtrsim 22 \kev$ which corresponds to
cosmic time $t \lesssim 3 \times 10^3$ sec. Namely after $t \simeq 3
\times 10^3$ sec, such a high energy proton can not be stopped in the
thermal bath, and it is inevitable to scatter off the ambient baryons
with the high energy.

As well as the high energy proton, we estimate the case of the high
energy neutron. The energy loss rate of the neutron through the
Coulomb scattering for $T \lesssim m_e$ is
\begin{eqnarray}
    \label{eq:dedt_n_c_rel}
    \frac{dE}{dt} = - \frac{3\pi \alpha^2 g_n^2 m_e}{m_n^2} n_e E^2.
\end{eqnarray}
The stopping time to lose the relativistic energy is 
\begin{eqnarray}
    \label{eq:stop_n_c_rel}
    \tau^n_{\rm stop} \simeq 1.68 \times 10^{-8}\sec x^{\frac32} e^x
    \left(\frac{E}{\gev}\right)^{-1}.
\end{eqnarray}
Here if we require $R^n_{\rm stop} \lesssim 1$, we find that the
temperature should be greater than 95 keV for the neutron stopping
which corresponds to the condition that cosmic time should be shorter
than 150 sec. In this case, after $t \simeq$ 150 sec, the high energy
neutron essentially inevitably scatters off the background baryons
before it stops. Under these situations, at longest, after $t \simeq$
150 sec the high energy nucleons necessarily scatter off the ambient
baryons through the strong interaction, and we would also have to
worry about the possibilities of the destruction of the light
elements. This means that the scattering process after $t \simeq$ 150
sec is beyond the limits of validity in our treatment in this
paper. For the problem, we will discuss it later.

As for $K^0_L$, it is never stopped in the electromagnetic plasma
because it does not interact with electrons and photons.  Therefore,
by using the energy dependent cross sections we will treat the
scattering off the ambient nucleons. To perform the computation, we
should know the correct energy distribution of $K^0_L$ produced
through the hadron fragmentation.

On the other hand,  for relatively longer lifetimes $\tau_{X}  \gtrsim
10^{4} \sec$, there is another interesting effects on BBN. The emitted
photons or charged leptons induce the electro-magnetic cascade
showers and produce many soft photons.~\footnote{
Even if the decay modes into hadrons are dominant ($B_h \sim
\order(1)$), the almost all parts of the energy of the parent particle
are transfered into photons and electrons because the hadrons decay
after they completely transfer their relativistic energy into the
thermal bath. In addition, it is expected that about 1/3 parts of the
produced hadrons are approximately $\pi^0$ and they decay as $\pi^0
\to \gamma \gamma$ with a much shorter lifetime ($\tau_{\pi^0} \simeq
10^{-16} \sec$).
}
Their spectrum has a cutoff at $E_\gamma^{\rm max}\simeq m_e^2/(22T)$.
If $E_\gamma^{\rm max}$ exceeds the binding energies of the light
elements, these photons destroy the light elements and change their
abundances~\cite{Holtmann:1999gd,Kawasaki:2001qr}. In fact, at $t \gtrsim 10^{4} (10^{6})
\sec$, the energy of the photon spectrum which are produced by the
decay of $X$, exceeds the deuterium ($^4$He) binding energy $B_2
\simeq 2.2$ ($B_4 \simeq 20$) MeV. However, because we are not
interested in the photodissociation here, we only study the case of
$\tau_{X} \lesssim 10^{4} \sec$. 

\subsection{Hadron Jets and collider experiments}
\label{subsec:jet_ex}

As an example of the hadronic decay, if the gravitino $\psi_{\mu}$ is
the parent particle $X$ whose mass is $m_X = \order$(1) TeV, it can
have net hadronic decay modes, e.g., $\psi_{\mu} \to \tilde{\gamma} q
\anti{q}$ ($q$: quark), with the branching ratio $B_h$. In this case,
$B_h$ can become $\sim {\cal O}$($\alpha$) at least even if the main
decay mode is only $\psi_{\mu} \to \tilde{\gamma} \gamma$
($\tilde{\gamma}$ : photino), because of the electromagnetic coupling
of the photon. As we quantitatively show later, about 1 hadrons are
produced for $B_h = 0.01$ and for the energy per two hadron jets,
$2E_{\rm jet} \sim 2/3m_X$, if we assume that the mechanism of the
hadron fragmentation is similar to the $e^+ e^-$ collider experiments.
In addition, the emitted high energy photon whose energy is about
$\sim m_X/2$ scatters off the background photon $\gamma_{\rm BG}$
and can produce a quark-antiquark pair.~\footnote{
Of course, there are some leptonic modes in the process, e.g. $\gamma
+ \gamma_{\rm BG} \to e^+ + e^-$. Thus, the net branching ratio into
hadrons is about $ \sim 60 \%$ in this energy.
}
Then, the center of mass energy is about $\sqrt{s} \sim $ 2 GeV and
produces about 3 hadrons which could effectively contribute to the
decay mode into hadrons as the branching ratio $B_h \sim
\order(0.01)$. Therefore, we should consider the hadronic decay modes
at least as $B_h = \order(0.01)$ in this case. On the other hand, if
the decay mode $\psi_{\mu} \to \tilde{g} g$ ($g$: gluon, and
$\tilde{g}$ : gluino) is kinematically allowed, $B_h$ may become close
to one.

For the other candidate of the parent particle, Polonyi field or
moduli, which appears in supergravity or superstring theory and has a
$\order(1) \tev$ mass, would also have a hadronic decay mode ($\phi
\to g g$ ).

\begin{figure}
      \begin{center}
          \centerline{\psfig{figure=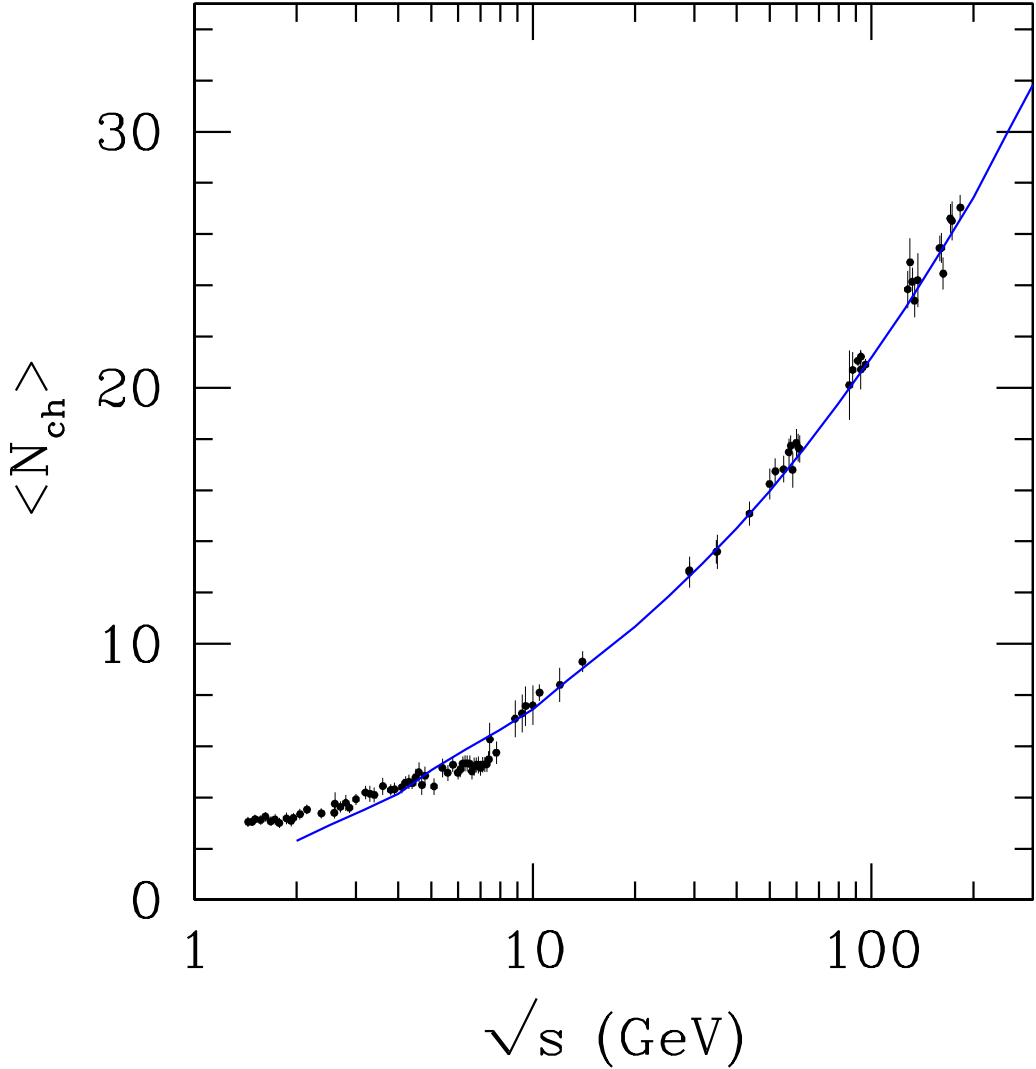,width=17cm}}
          \caption{
          Plot of the averaged charged-particle multiplicity $\langle
          N_{\rm ch}\rangle$. This represents the total number of the
          charged hadrons emitted per $e^+ e^-$ annihilation and per
          two hadron jets as a function of $\sqrt{s}$ (= 2 $E_{\rm
          jet}$), where $\sqrt{s}$ denotes the center of mass energy,
          and $E_{\rm jet}$ is the energy per one hadron jet. The
          solid line denotes the value obtained by using the JETSET
          7.4 Monte Carlo event generator. The filled circle denotes
          the data points of $e^+ e^-$ collider experiments. Error is
          quadratically added for the statistical and systematic
          one. Here $\langle N_{\rm ch}\rangle$ is defined as the
          value after both $K_S$ and $\Lambda^0$ had completely
          finished to decay.
          }
          \label{fig:nch}
      \end{center}
\end{figure}

Fortunately, we can estimate the number and energy distribution of the
produced hadrons by using the JETSET 7.4 Monte Carlo event
generator~\cite{Sjostrand:1994yb}. This FORTRAN package computes the hadron
fragmentation for the $q \anti{q}$ event ($q$: quark) in the $e^+ e^-$
annihilation and predicts the energy distribution of the products to
agree with the $e^+ e^-$ collider experiments. In Fig.~\ref{fig:nch}
we plot the averaged charged-particle multiplicity $\langle N_{\rm
ch}\rangle$ which represents the total number of the charged hadrons
emitted per $e^+ e^-$ annihilation and per two hadron jets as a
function of $\sqrt{s}$ (= 2 $E_{\rm jet}$).~\footnote{
Here $\langle N_{\rm ch}\rangle$ is defined as the value after both
$K_S$ and $\Lambda^0$ had completely finished to decay, where their
lifetimes are $\tau_{K^0_S} = 0.89 \times 10^{-10}$ sec and
$\tau_{\Lambda^0} = 2.63 \times 10^{-10}$ sec respectively. As we
showed in section~\ref{subsec:timescale}, we should not treat any
particles with the shorter lifetime than $\sim 10^{-8}$ sec in this
situation.
}
Recently LEPII experiments (ALEPH, DELPHI, L3 and OPAL) give us the
useful data for $\sqrt{s}$ = 130 -- 183 GeV.  Therefore, now a number
of experimental data are available at least up to $\sqrt{s} \simeq$
183 GeV~\cite{PDG}. The filled circle denotes the data points of $e^+
e^-$ collider experiments. From Fig.~\ref{fig:nch} we find that the
predicted $\langle N_{\rm ch}\rangle$ excellently agrees with the
experimental values. Thus, in this situation we use the JETSET 7.4 to
infer the spectrum of the emitted hadrons extrapolating to the various
higher energies.

\begin{figure}
      \begin{center}
          \centerline{\psfig{figure=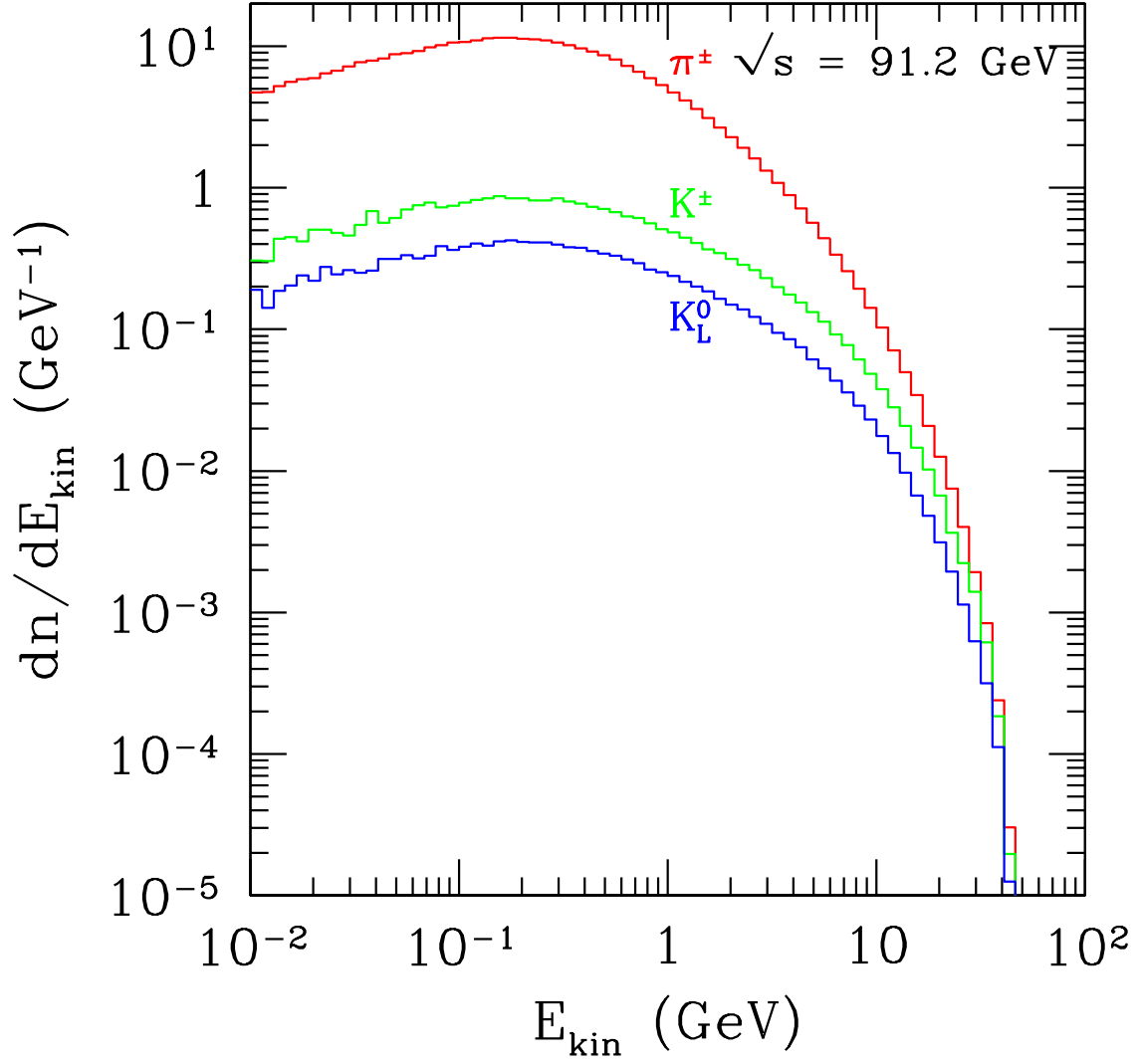,width=17cm}}
          \caption{
          Plot of the spectrum of the produced mesons ($\pi^+$ +
          $\pi^-$, $K^+$ + $K^-$, and $K^0_L$) as a function of the
          kinetic energy $E_{\rm kin}$. This is the case that the
          center of mass energy is $\sqrt{s} = 91.2$ GeV which
          corresponds to the $Z^0$ resonance. They are computed by
          using the JETSET 7.4 Monte Carlo event generator.
          }
          \label{fig:hadron_spec}
      \end{center}
\end{figure}

\begin{figure}
      \begin{center}
          \centerline{\psfig{figure=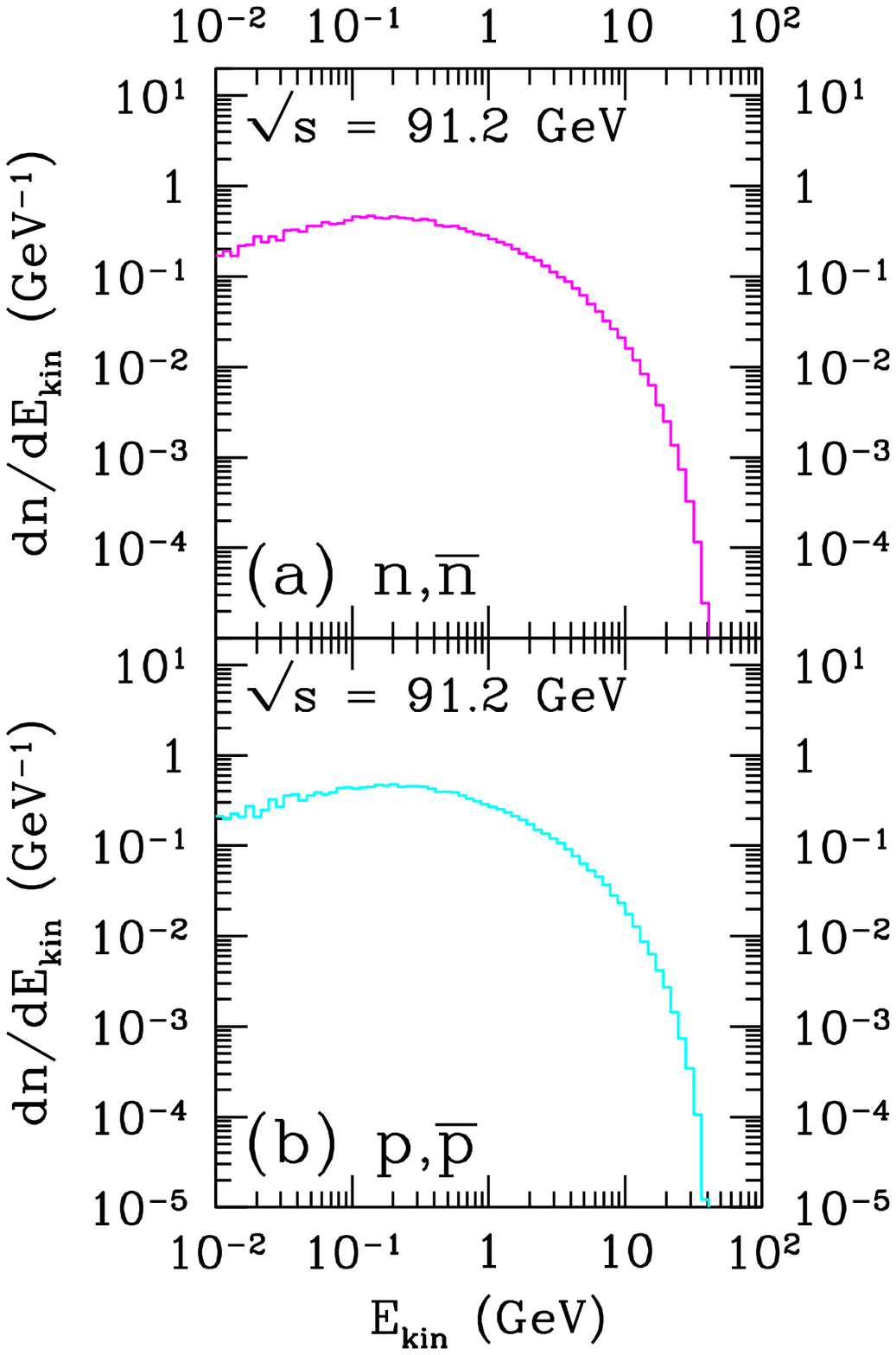,width=17cm}}
          \caption{
          Plot of the spectrum of the produced baryons ((a) $n +
          \anti{n}$ and (b) $p + \anti{p}$) as a function of the
          kinetic energy $E_{\rm kin}$. This is the case that the
          center of mass energy is $\sqrt{s} = 91.2$ GeV which
          corresponds to the $Z^0$ resonance. They are computed by
          using the JETSET 7.4 Monte Carlo event generator.
          }
          \label{fig:baryon_spec}
      \end{center}
\end{figure}

\begin{figure}
      \begin{center}
          \centerline{\psfig{figure=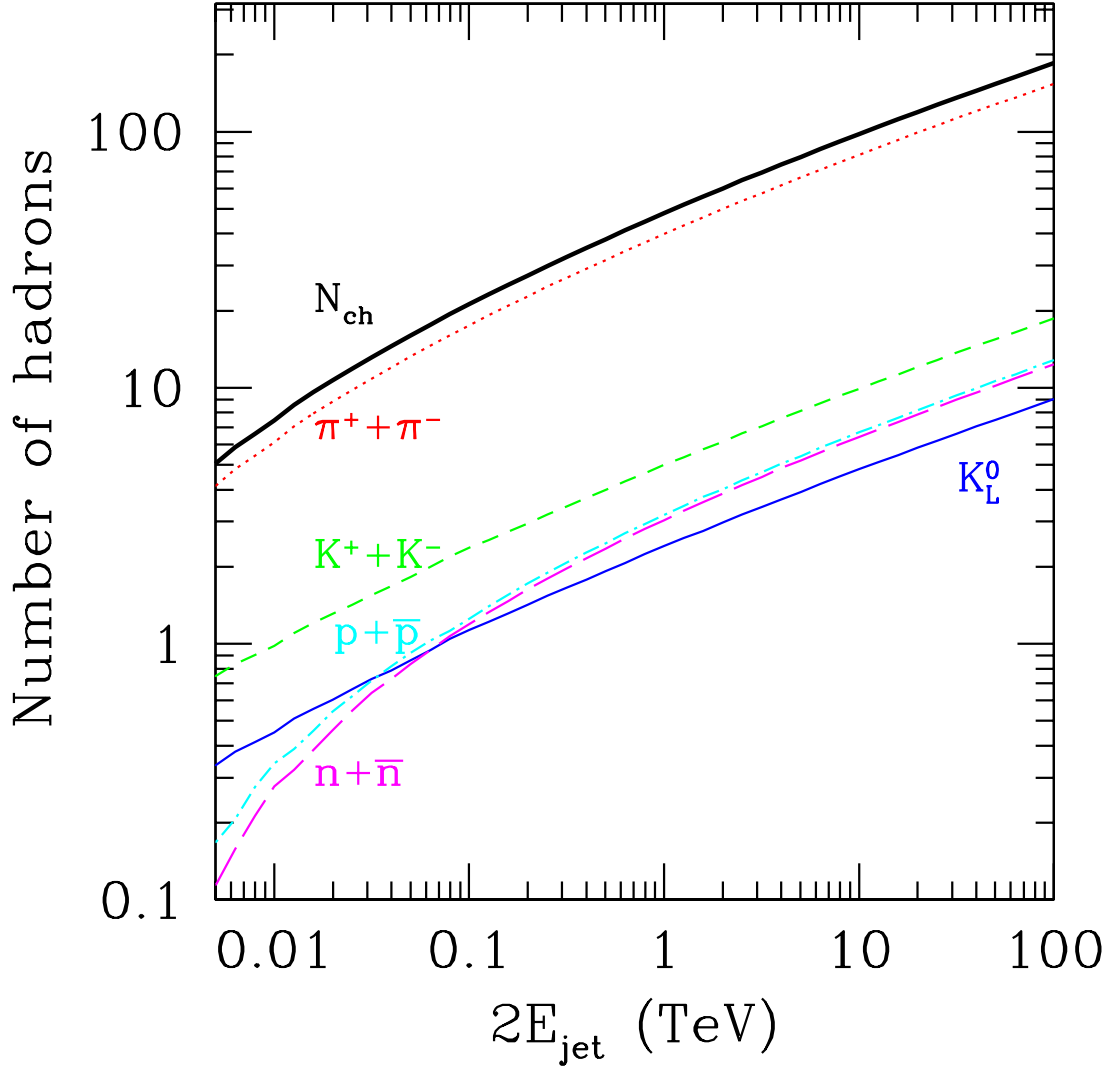,width=17cm}}
          \caption{
          Plot of the averaged number of the produced hadrons as a
          function of $2 E_{\rm jet} (=\sqrt{s})$, where $E_{\rm jet}$
          denotes the energy of one hadron jet. The number is defined
          by the value per two hadron jets. $\langle N_{\rm
          ch}\rangle$ denotes the averaged charged-particle
          multiplicity (thick solid line). The number is obtained by
          summing up the energy distribution. The dotted line is
          $\pi^+ + \pi^-$, the short dashed line is $K^+ +K^-$, the
          thin solid line is $K^0_L$, the dot-dashed line is $p +
          \anti{p}$, and the long dashed line is $n + \anti{n}$. They
          are computed by using the JETSET 7.4 Monte Carlo event
          generator.
          }
          \label{fig:nch_large}
      \end{center}
\end{figure}

In Fig.\ref{fig:hadron_spec} we plot the spectrum of the produced
mesons ($\pi^+$ + $\pi^-$, $K^+$ + $K^-$, and $K^0_L$) as a function
of the kinetic energy $E_{\rm kin}$. This is the case that the center
of mass energy is $\sqrt{s} = 91.2$ GeV which corresponds to the $Z^0$
resonance. In similar fashion, in Fig.~\ref{fig:baryon_spec} we plot
the spectrum of the produced baryons ((a) $n + \anti{n}$, and (b) $p +
\anti{p}$).  In Fig.~\ref{fig:nch_large} we plot the averaged number
of the produced hadron per two hadron jets as a function of $2 E_{\rm
jet}$, which is obtained by summing up the energy distribution. From
Fig.~\ref{fig:nch_large}, we find that almost all hadrons are composed
of pions.

\subsection{Cross sections of hadron-nucleon scattering}
\label{subsec:cross-section}

Because in this paper we are interested in the BBN epoch, i.e. $T
\lesssim \order(1)$ MeV, the temperature is much less than the typical
mass of the emitted hadrons,
e.g. $m_{H_i}~=$~$\order$(100)\mev~--~$\order$(1) $\gev$. As we
discussed in section~\ref{subsec:stopping}, as long as the temperature
is relatively high enough ($T \gtrsim 95 \kev$), the emitted high
energy hadrons ($\pi^{\pm}, K^{\pm}$, $p$, and $n$) have completely
lost their relativistic energies through the electromagnetic
interaction in the thermal plasma and are quickly thermalized except
for neutral kaon $K_L^0$. Then only the exothermic process is relevant
for the hadron to scatter off the background baryons through the
strong interaction because it has just a little kinetic energy of the
order of the temperature $T$. Of course, such a low energy hadron can
not destroy the background $\hefour$. Concerning exothermic reactions,
it is well-known that the cross section $\sigma$ is nearly inversely
proportional to the velocity $v$ of the projectile particle in the low
energy. Namely $\sigma v$ almost does not have a $v$ dependence and is
nearly a constant for the beam energy. Except for $K_L^0$, therefore,
we can use the threshold cross section instead of the thermal-averaged
cross section. Here we adopt the results of the thermal-averaged cross
section in Ref.~\cite{Reno:1988qw}.

The thermally averaged cross sections for $\pi^{\pm}$ are
\begin{eqnarray}
    \label{eq:sigma_pi}
    {}&& \langle\sigma v\rangle^{\pi^+}_{n \rightarrow p} = 1.7 \ \mb, \\ 
    {}&& \langle\sigma v\rangle^{\pi^-}_{p \rightarrow n} =
    1.5C_{\pi}(T) \ \mb,
\end{eqnarray}
where $C_{H_i}(T)$ is the Coulomb correction factor when the beam
particle $H_i$ is the charged one. Because the reaction $p^+ + \pi^-
\rightarrow n + \cdot \cdot \cdot$ is enhanced due to the
opposite-sign charge of the initial state particles, we should correct
the strong interaction rates by simply multiplying $C_{H_i}(T)$ to
that which are obtained by ignoring the Coulomb corrections. The
Coulomb correction factor is generally estimated by
\begin{equation}
    \label{eq:coulomb}
    C_{H_i}(T) = \frac{2\pi \xi_i(T)}{1 - e^{-2\pi \xi_i(T)}},
\end{equation}
where $\xi_i(T) = \alpha \sqrt{\mu_i/ 2T}$, $\alpha$ is the fine
structure constant, and $\mu_i$ is the reduced mass of the hadron
$H_i$ and the nucleon. 

The thermally-averaged cross sections for $K^-$ are 
\begin{eqnarray}
    \label{eq:sigma_ka-}
    {}&& \langle\sigma v\rangle^{K^-}_{n \rightarrow p} = 26 \ \mb, \\ 
    {}&& \langle\sigma v\rangle^{K^-}_{n \rightarrow n} = 34 \ \mb, \\ 
    {}&& \langle\sigma v\rangle^{K^-}_{p \rightarrow n} = 31  C_{K}(T)
      \ \mb, \\ 
    {}&& \langle\sigma v\rangle^{K^-}_{p \rightarrow p} = 14.5
    C_{K}(T) \ \mb.
\end{eqnarray}
Here we ignore $K^+$ interaction because $n + K^+ \rightarrow p + K^0$
is the endothermic reaction which has $Q = 2.8$ MeV, and it is
expected that the kinetic energy of $K^+$ is less than $Q$.

\begin{figure}
      \begin{center}
          \centerline{\psfig{figure=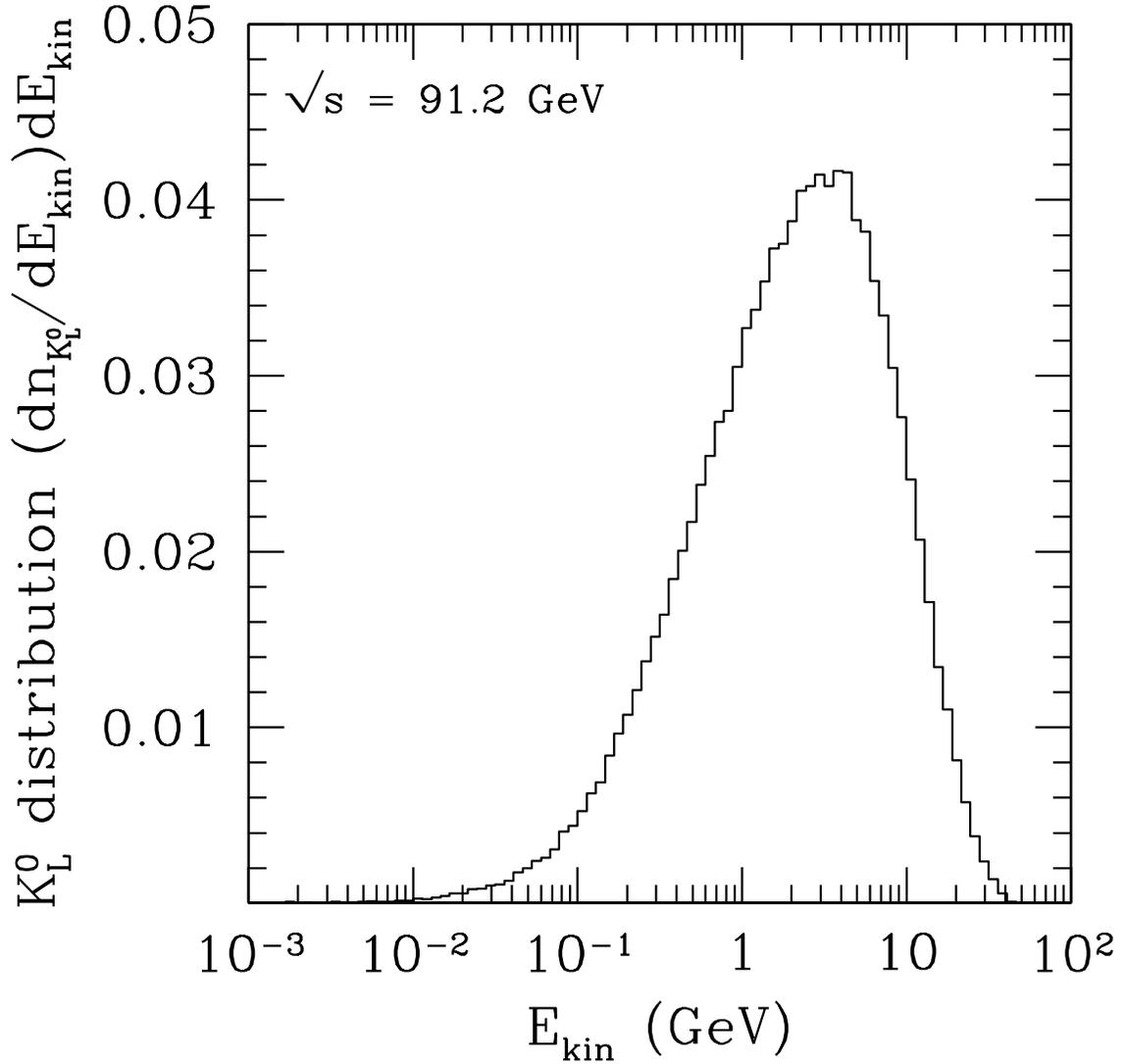,width=17cm}}
          \caption{
          Plot of the distribution of $K^0_L$ produced in the $e^+
          e^-$ annihilation as a function of the kinetic energy. It is
          the case that the center of mass energy is $\sqrt{s} = 91.2$
          GeV which corresponds to the $Z^0$ resonance. It is computed
          by the JETSET 7.4 Monte Carlo event generator.
          }
          \label{fig:kl_dist}
      \end{center}
\end{figure}

As for neutral kaon $K^0_L$, there are no adequate experimental data
of the differential cross sections as a function of the beam energy to
use in our current purpose. It is very serious for us because $K^0_L$
does not lose their relativistic energy and is never stopped in the
thermal bath, and then we should know the differential cross sections
in whole relevant energy range. For example, in Fig.~\ref{fig:kl_dist}
we find that the source distribution function of $K^0_L$ is spread in
the wide energy range. At least we want to obtain the data of the
cross sections for the typical $K^0_L$-beam energy, $E_{\rm beam}$ =
10 MeV -- 1 TeV, where $E_{\rm beam}$ is the kinetic energy of
$K^0_L$. In this situation, we should estimate the data table of the
cross sections of the $K^0_L$ scattering by using the other existing
experimental informations.

Here we assume that $K^0_L$ scatters off the nucleon $N$ as a
combination of 1/2 $K^0$ and 1/2 $\anti{K}^0$ because in fact $K^0_L$
is nearly the linear combination of $K^0$ and $\anti{K}^0$ states that
$|K^0_L\rangle \simeq (|K^0\rangle -
|\anti{K}^0\rangle)/\sqrt{2}$.~\footnote{
Of course, the CP violation effect does not change our rough estimates
at all and is not important here.
}
In addition, we assume that the strangeness of $K^0$ ($\anti{K}^0$) is
similar to $K^+$ ($K^-$) because $K^0 = d\anti{s}$ ($\anti{K}^0 =
s\anti{d}$) contains $\anti{s}$ ($s$) ($s$: strange quark, and $d$:
down quark). Of course, the above assumption is not wrong very much
under the isospin SU(2) transformation for the $u\choose d$ doublet
($u$: up quark) because we cannot imagine that there exists a special
coherent interference in the inelastic scattering.

In this assumption, we would also have to worry about the effect of
the Coulomb corrections because actually $K^0 N$ scatterings are not
supposed to suffer from any electric charges. From
Eq.~(\ref{eq:coulomb}), however, we find that the Coulomb correction
is less than 10$\%$ at most in both cases of the attractive force and
the repulsive one as long as the kinetic energy of $K^{\pm}$ is more
than $\order$(10) MeV. Therefore, we can ignore the Coulomb correction
and the above assumption would be reasonable in this situation.

\begin{figure}
      \begin{center}
          \centerline{\psfig{figure=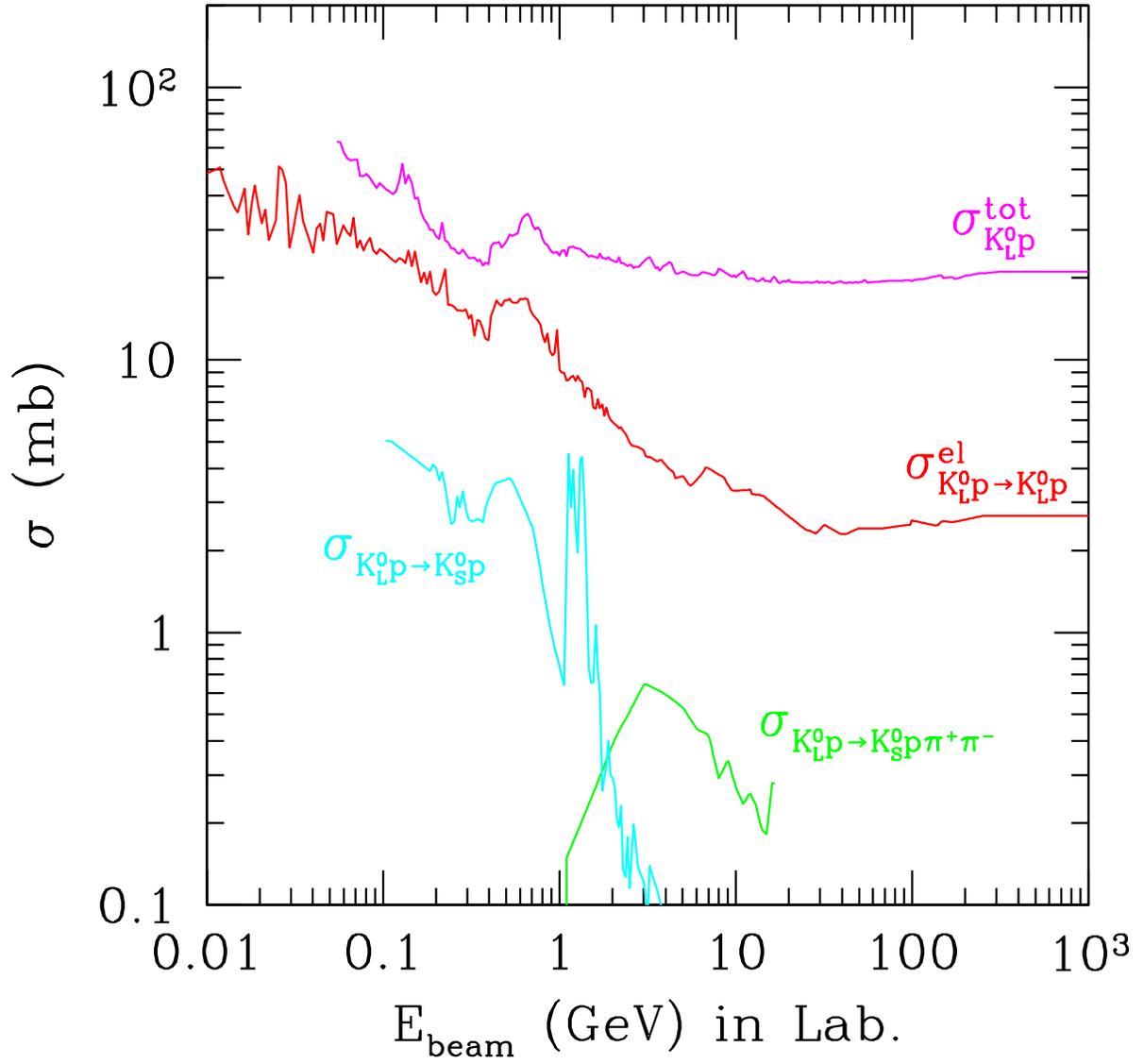,width=17cm}}
          \caption{
          Plot of the data of the cross sections as a function of the
          kinetic energy of the $K_L^0$ beam.
          }
          \label{fig:kl_ebeam}
      \end{center}
\end{figure}

Fortunately, we have good compilations of the experiments for the
total cross section and the elastic cross section for $K^{+}p$ and
$K^{-}p$~\cite{PDG}.  Thus, by averaging them we can estimate the
total $\sigma^{\rm tot}_{K^0 p}$ and elastic cross section
$\sigma^{\rm el}_{K^0_L p}$ respectively. In Fig.~\ref{fig:kl_ebeam},
we plot the obtained total and elastic cross sections for $K^0_L p$
scattering. It is happy that the obtained total cross sections agree
with the direct experimental data and the theoretical predictions
marginally within a few tens percent although they were studied only
in the low energy regions for $E_{\rm beam}\lesssim 0.7$
GeV~\cite{exp_K_L}. In addition, we have the experimental data of the
inelastic scatterings, $K^0_Lp \to K^0_Sp$ and $K^0_Lp \to
K^0_Sp\pi^+\pi^-$~\cite{durham} which are also plotted in
Fig.~\ref{fig:kl_ebeam}. Now we assume that the cross section of the
inter-converting reaction $K^0_L + p \to n + \cdot \cdot \cdot$ is
obtained by $\sigma^{K^0_L}_{p \to n} = 1/2[\sigma^{\rm tot}_{K^0_Lp} -
(\sigma^{\rm el}_{K^0_Lp \to K^0_Lp}+\sigma_{K^0_Lp \to
K^0_Sp}+\sigma_{K^0_Lp \to K^0_Sp\pi^+\pi^-})]$ because the final
states of the inelastic scattering without $K^0_L p \to K^0_S p +
\tenten$ are $K N \pi$, $\Lambda^0 \pi$, or $\Sigma \pi$, and it is
approximately expected
that either $p$ or $n$ appears in a closely even probability.~\footnote{
The branching ratios are presented as $\Lambda^0 \to n\pi^0 (35\%)$,
$p\pi^-(63.9\%)$; $\Sigma^0 \to \Lambda^0\gamma (100\%)$; $\Sigma^+
\to n\pi^+(48.3\%)$, $p\pi^0(51.6\%)$; $\Sigma^- \to
n\pi^-(99.9\%)$~\cite{PDG}. We also ignore the multiple production
process of baryons because the center of mass energy is too low for
the process to dominate the other reactions.
}
Then, we get the remaining cross section as $\sigma^{K^0_L}_{p \to p}
= \sigma^{\rm tot}_{K^0_Lp} - \sigma^{K^0_L}_{p \to n}$. 

About neutron-$K^0_L$ scattering, we could have performed the similar
treatments.  However, compared to the cases of proton, we do not have
adequate compilations for the neutron-$K^{\pm}$ process. On the other
hand, the data tell us that we can approximately regard them as the
cross sections of the proton-$K$ scattering within a few tens percent
in the high beam energies ($E_{K} \gtrsim 100$ MeV). The theoretical
reason is that the strong interaction does not distinguish between
proton and neutron in such a high energy. Under these circumstances,
we assume that the cross section of $K^0_L n$ is as same as $K^0_L p$
with a few tens percent error.

\begin{figure}
      \begin{center}
          \centerline{\psfig{figure=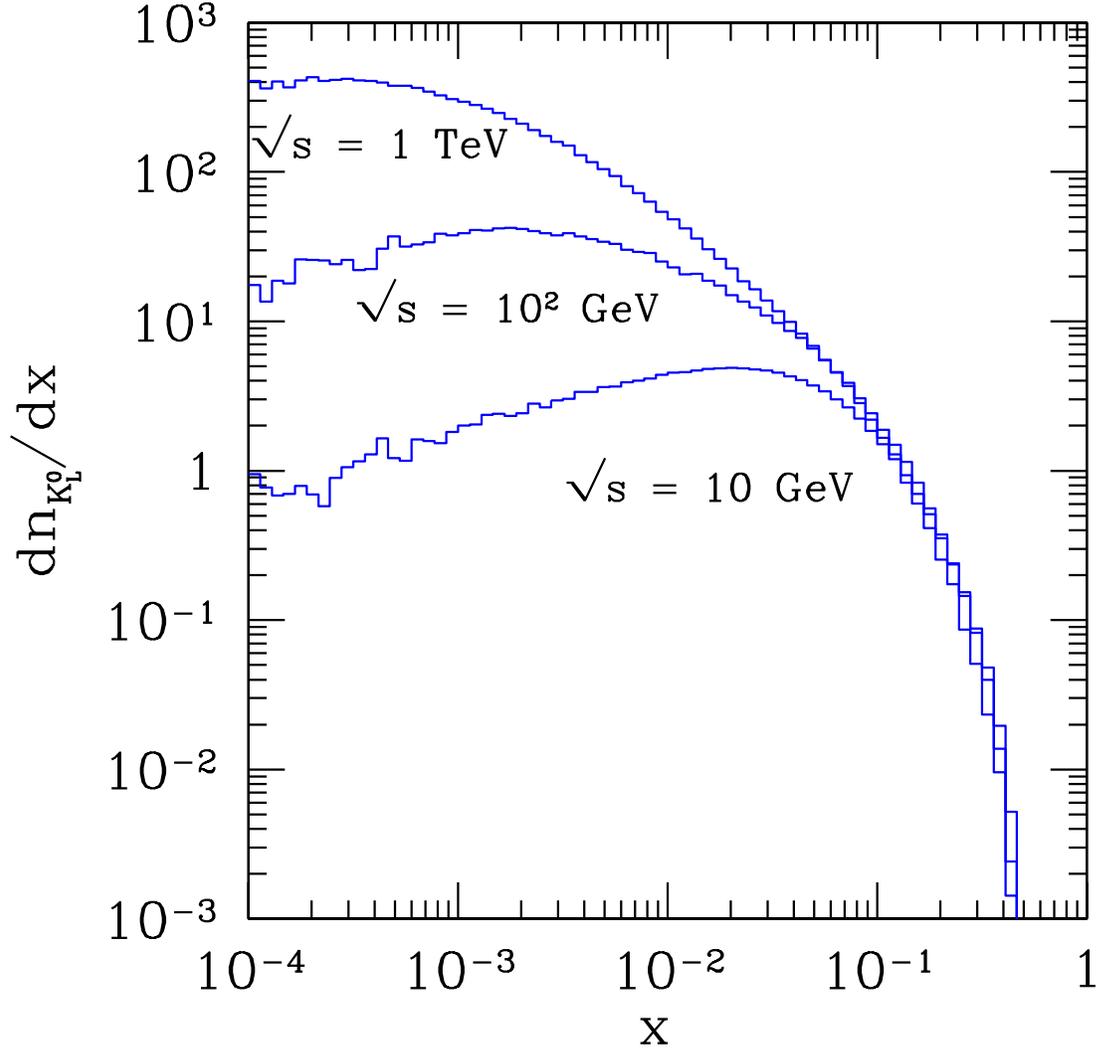,width=17cm}}
          \caption{
          Plot of the spectrum of the $K^0_L$ produced through the
          hadron fragmentation of $q\anti{q}$ pair emitted from $e^+
          e^-$ annihilation. $x$ $(\equiv E_{\rm kin}/\sqrt{s})$
          denotes the normalized kinetic energy $E_{\rm kin}$, and
          $\sqrt{s}$ denotes the center of mass energy of $e^+ e^-$
          collision. They are computed by using the JETSET 7.4 Monte
          Carlo event generator.
          }
          \label{fig:dndx}
      \end{center}
\end{figure}

\begin{figure}
      \begin{center}
          \centerline{\psfig{figure=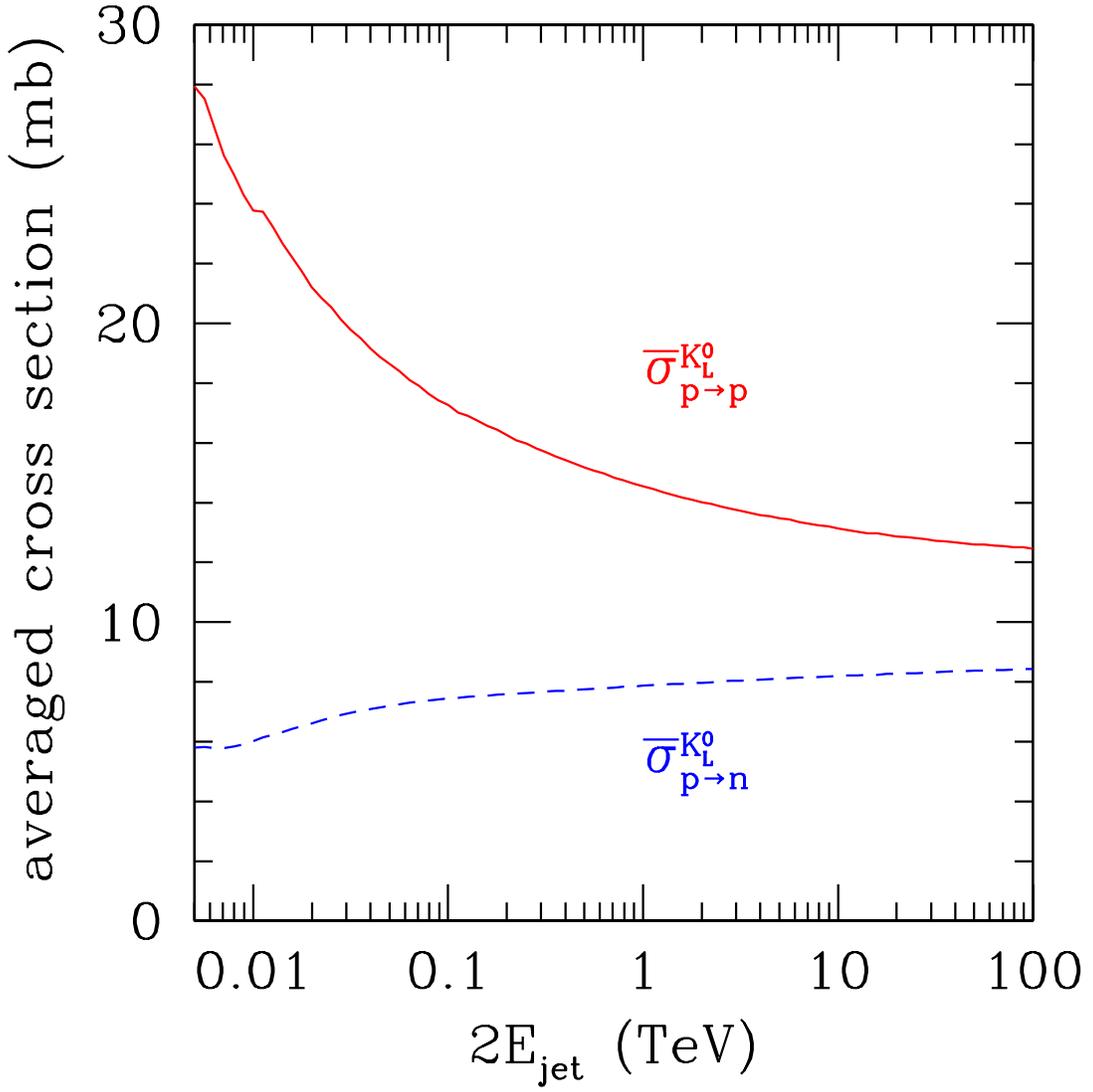,width=17cm}}
          \caption{
          Plot of the averaged cross sections for $p+K^0_L \to p+
          \tenten$ and $p+K^0_L \to n +\tenten$ as a function of the
          energy of two jets (= $2E_{\rm jet}$). 
          }
          \label{fig:sigave}
      \end{center}
\end{figure}

To perform the numerical computations including the $K^0_L$-injection
effects in BBN, it is useful to average the cross sections by the
energy spectrum of $K^0_L$. As we discussed in the previous
subsection, we can use the JETSET 7.4 Monte Carlo generator and get
the energy spectrum of emitted $K^0_L$ in wide range of the source
energy. For example, we can see the spectrum of the produced $K^0_L$
for various energies in Fig.~\ref{fig:dndx}. Then, we get the averaged
cross sections, $\anti{\sigma}^{K^0_L}_{p \to p}$ and
$\anti{\sigma}^{K^0_L}_{p \to n}$, as the convolutions of the data of
the cross sections with the energy spectrum of $K^0_L$
(Fig.~\ref{fig:sigave}).

Concerning the emitted nucleons, we basically follow the Reno and
Seckel's treatment that we regard the nucleon-antinucleon pair as a
kind of a meson $N\anti{N}$~\cite{Reno:1988qw}. Then, the $N\anti{N}$ meson
induces the inter-conversion $N + N\anti{N} \rightarrow N' + \cdot
\cdot \cdot$. In Ref.~\cite{Reno:1988qw}, we have the thermally-averaged cross
sections,
\begin{eqnarray}
    \label{eq:sigma_nuc}
    {}&& \langle\sigma v\rangle^{n\anti{n}}_{n \rightarrow n} =  37 \ \mb, \\ 
    {}&&  \langle\sigma v\rangle^{n\anti{n}}_{p \rightarrow n} =  28 \ \mb, \\ 
    {}&& \langle\sigma v\rangle^{p\anti{p}}_{n \rightarrow p} =  28 \ \mb, \\ 
    {}&&  \langle\sigma v\rangle^{p\anti{p}}_{p \rightarrow p} =  37 \ \mb. 
\end{eqnarray}
As we discussed in the previous sections, however, the late time
emission of the high energy nucleons would induce the destruction of
light elements for $T \lesssim 95 \kev$. However, for the moment we
treat the nucleons as if they are approximately thermalized. We will
also discuss the modification on the result caused by the above simple
assumption later.

\subsection{Formulation in hadron-injection scenario}
\label{subsec:formulation}

We formulate the time evolution equations in the late-time
hadron-injection scenario here.  As we have mentioned in the previous
subsections, the hadron injection at the beginning of BBN enhances the
inter-converting interactions between neutron and proton equally and
the freeze out value of $n/p$ is extremely increased. Then the time
evolution equations for the number density of a nucleon $N (=p, n)$ is
represented by
\begin{equation}
     \label{eq:difeqN}
     \frac{dn_N}{dt} + 3 H(t) n_N =
     \left[\frac{dn_N}{dt}\right]_{\rm weak} - B_h \Gamma_X n_X \left(
       K_{N \rightarrow N'} - K_{N' \rightarrow N} \right),
\end{equation}
where $H(t)$ is Hubble expansion rate, $[dn_N/dt]_{\rm weak}$ denotes the
contribution from the usual weak interaction rates as well as SBBN,
$B_h$ is the branching ratio of the hadronic decay mode of $X$, $n_{X}$
is the number density of $X$, $K_{N \rightarrow N'}$ denotes the
average number of the transition $N \rightarrow N'$ per one $X$ decay.

The average number of the transition $N \rightarrow N'$ per one $X$
decay is expressed by
\begin{equation}
     \label{eq:Knn}
     K_{N \rightarrow N'} = \sum_{H_i} \frac{N_{\rm
     jet}}{2}N^{H_i}R^{H_i}_{N \rightarrow N'},
\end{equation}
where $H_i$ runs the hadron species which are relevant to the nucleon
inter-converting reactions, $N_{\rm jet}$ is the number of the hadron
jet per one X decay, $N^{H_i}$ denotes the average number of the
emitted hadron species $H_i$ per one $X$ decay. $N^{H_i}$ is presented
in Fig.~\ref{fig:nch_large} as a function of $2 E_{\rm jet}$, where
$E_{\rm jet}$ is the energy of a hadron jet.  $R^{H_i}_{N \rightarrow
N'}$ denotes the probability that a hadron species $H_i$ induces the
nucleon transition $N \rightarrow N'$ and is represented by
\begin{equation}
     \label{eq:trans_prob}
     R^{H_i}_{N \rightarrow N'} =
      \frac{\Gamma^{H_i}_{N \rightarrow N'}}{\Gamma^{H_i}_{\rm dec} +
      \Gamma^{H_i}_{\rm abs}},
\end{equation}
where $\Gamma^{H_i}_{\rm dec} = \tau_{H_i}^{-1}$ is the decay rate of
$H_i$, $\tau_{H_i}$ is the lifetime, and $\Gamma^{H_i}_{\rm abs} \equiv
\Gamma^{H_i}_{N \rightarrow N'}+\Gamma^{H_i}_{N' \rightarrow
N}+\Gamma^{H_i}_{N \rightarrow N}+\Gamma^{H_i}_{N' \rightarrow N'}$ is
the total absorption rate of $H_i$.

\begin{figure}
      \begin{center}
          \centerline{\psfig{figure=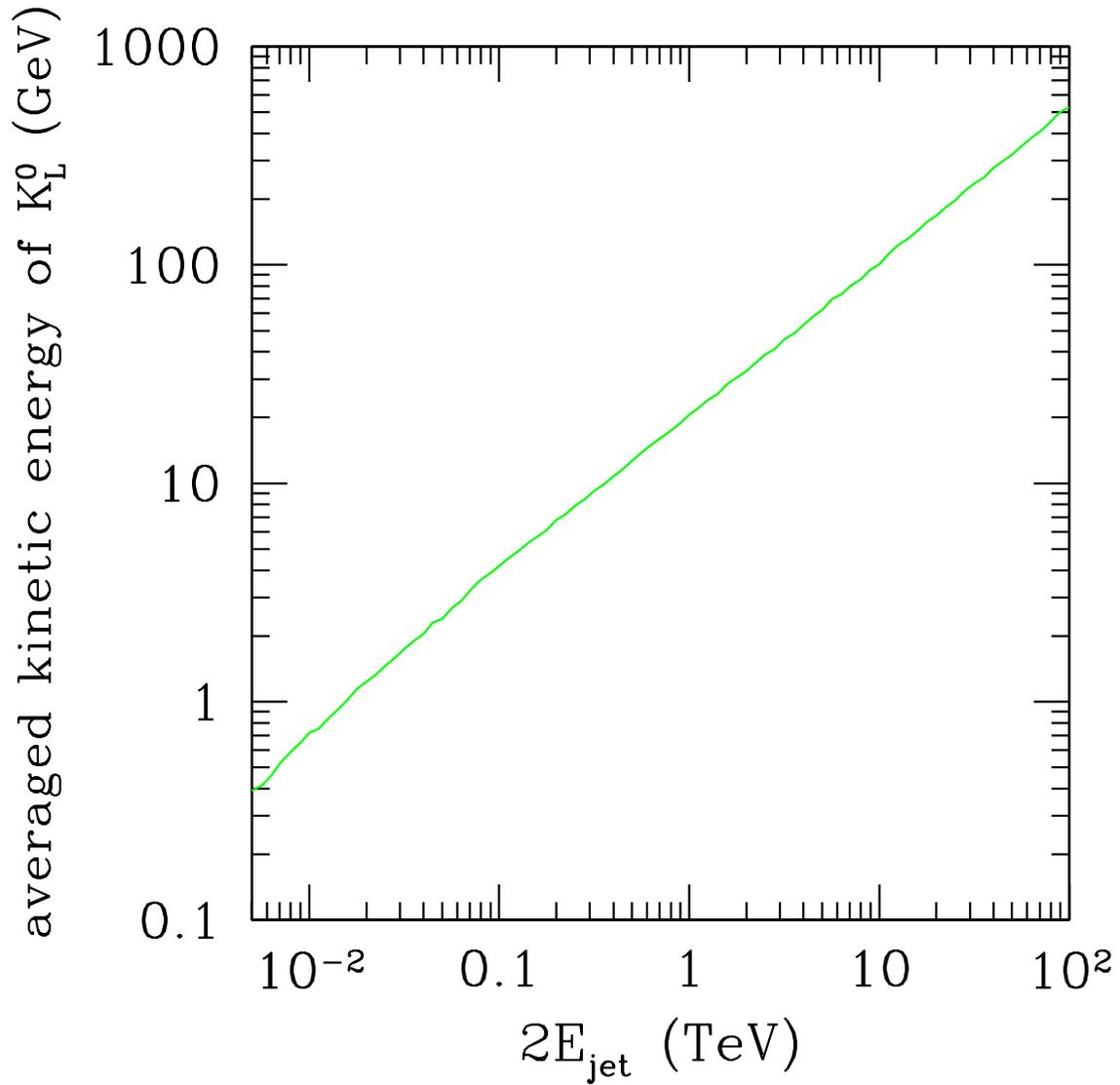,width=17cm}}
          \caption{
          Plot of the mean kinetic energy of $K_L^0$ which is obtained
          by weighting the kinetic energies for their distribution as
          a function of $2 E_{\rm jet}$, where $2E_{\rm jet}$ is the
          energy of two hadron jets.
          }
          \label{fig:kinave}
      \end{center}
\end{figure}

Because the emitted high energy $K_L^0$ is not stopped in the thermal
bath, its lifetime becomes longer by a factor of $E_{K_L^0}/m_{K_L^0}$
due to the relativistic effect. Then, the decay rate is estimated by
$\Gamma^{K_L^0}_{\rm dec} = \tau^{-1}_{K_L^0}
m_{K_L^0}/E_{K_L^0}$. Because the emitted $K_L^0$'s are distributed in
the wide energy range, for convenience we computes the mean kinetic
energy $\anti{E}_{K_L^0}$ which is obtained by weighting the kinetic
energies for their distribution (see Fig.~\ref{fig:dndx}).  In
Fig.~\ref{fig:kinave}, $\anti{E}_{K_L^0}$ is plotted as a function of
$2 E_{\rm jet}$.

\section{BBN computation in hadron-injection scenario
and comparison with observations}
\label{sec:comparison}


In this section we perform the BBN computations in the
hadron-injection scenario. Then we compare the theoretical prediction
of the light element abundances with the observational light element
abundances. In the computations we assume that the massive particle
$X$ decays into three bodies ($E_{\rm jet}=m_{X}/3$) and two jets are
produced at the parton level (i.e. the number of jets $N_{\rm
jet}=2$).  The above choice of a set of model parameters $E_{\rm jet}$
and $N_{\rm jet}$ is not unique in general and is obviously model
dependent. For $E_{\rm jet}$ however, since we study the wide range of
the mass, we can read off the results by rescaling the mass parameter.
In addition, for the modification of $N_{\rm jet}$ since the second
term in the right hand side in Eq.~(\ref{eq:difeqN}) scales as
$\propto N_{\rm jet}$, we only translate the obtained results
according to the above scaling rule and push the responsibility off
onto the number density $n_{X}$.

As we noted in the previous sections, it is a remarkable feature that
the predicted $\hefour$ mass fraction $Y$ tends to increase in the
hadron injection scenario because $\hefour$ is the most sensitive to
the freeze-out value of the neutron to proton ratio in the beginning
of BBN.  Since protons which are more abundant than neutrons are
changed into neutrons through the strong interactions rapidly, the
freeze out value of $n/p$ increase extremely if once the net hadrons
are emitted.  In addition, D is also sensitive to the neutron number
after $T \lesssim 0.1 \mev$ because the free neutrons can not burn
into $\hefour$.

\begin{figure}
      \begin{center}
          \centerline{\psfig{figure=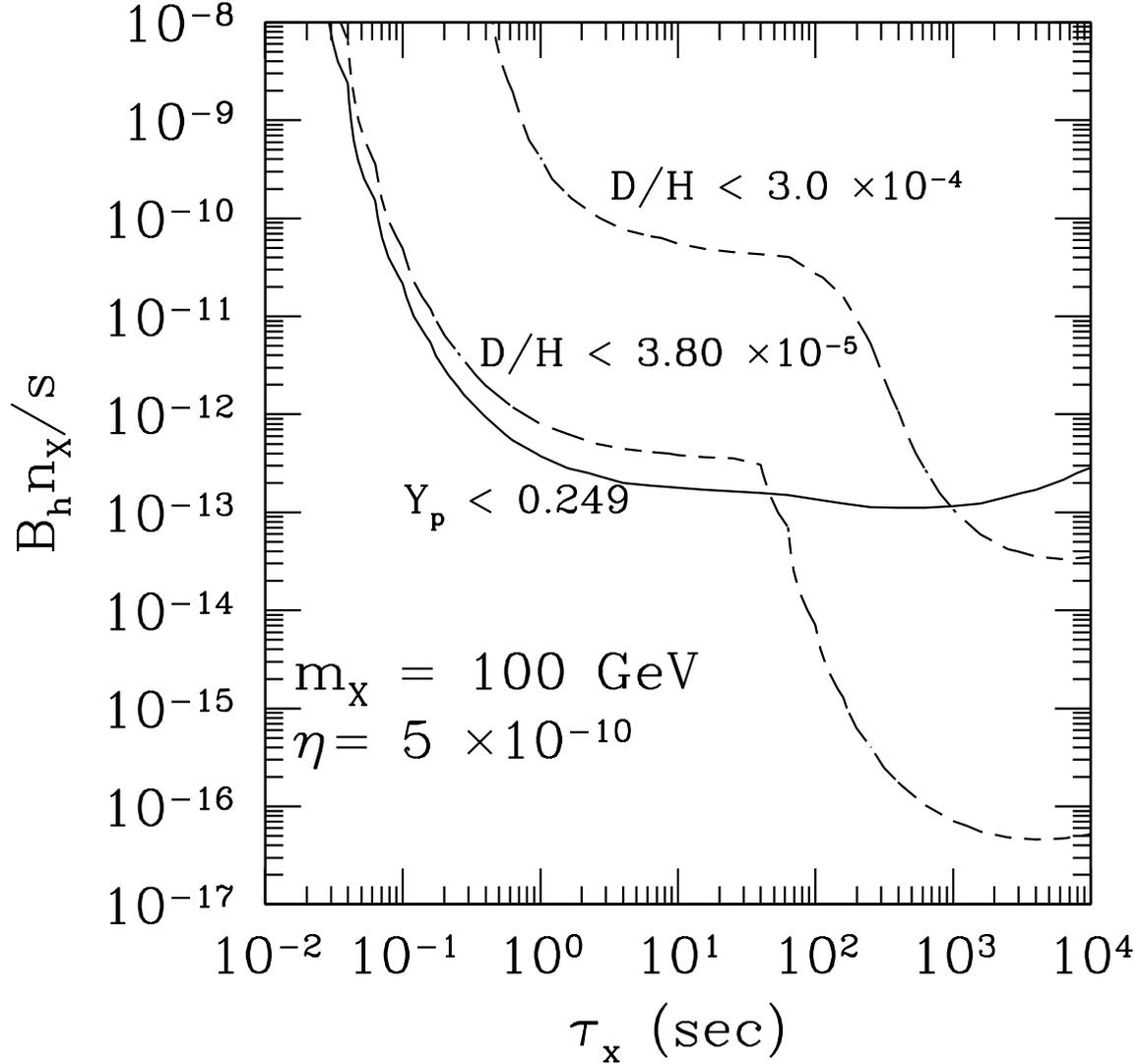
          ,width=17cm}}
          \caption{
          Plot of the rough upper bound of $B_h n_X /s$ from the
          observational $2\sigma$ upper bounds of $^4$He (solid line),
          and D (dashed line) for high D or low D as a function of the
          lifetime of the massive particle $X$. $B_h$ is the hadronic
          branching ratio of $X$, and $n_X /s$ denotes the number
          density of $X$ per entropy density $s$. Here the baryon to
          photon ratio is $\eta = 5 \times 10^{-10}$ and the mass of
          $X$ ($m_X$) is fixed to be 100 GeV. The observational upper
          bounds are obtained by adding the errors in quadrature.
          }
          \label{fig:non_MC}
      \end{center}
\end{figure}

To see the rough tendency, we plot the upper bounds for $B_h n_X /s$ in
Fig.~\ref{fig:non_MC} which come from each observational $2\sigma$
upper bound for $^4$He, and D as a function of the lifetime $\tau_X$
at the baryon
to photon ratio $\eta = 5 \times 10^{-10}$.~\footnote{
The $\liseven$ abundance is mildly constrained from the observation
and is much weaker than the others. In addition since it has a
complicated dependence for $\eta$, we do not plot it here. Of course,
however, we include $\liseven$ in Monte Carlo simulation and Maximum
Likelihood analysis which will be discussed below.
}
$B_h$ is the hadronic branching ratio of $X$, and $n_X /s$ denotes the
number density of $X$ per entropy density $s$. The mass is fixed to be
typical value, $m_X = 100 \gev$.  From the figure, we find that for
the shorter lifetime $\tau_X \lesssim 10^{-2} \sec$, the hadron
injections do not affect the freeze-out value of $n/p$ and do not
change any predictions of SBBN.  However, if the lifetime becomes
longer $\tau_{X} \gtrsim 10^{-2} \sec$, the freeze-out value of $n/p$
ratio is increased by the hadron-induced inter-converting interactions
and the produced neutron increases the $^4$He abundance because most
of the free neutrons burn into $^4$He through D. Then, $n_X/s$ is
strongly constrained by the upper bound of the observational $^4$He
abundance.  For $\tau_{X} \gtrsim 10^{2} \sec$, since the produced
free D can no longer burn into $^4$He, the extra free neutrons still
remain in D. Then $n_X/s$ is severely constrained by the upper bound
of the observational D/H. For the constraint from high D, i.e. D/H $<
3.0 \times 10^{-4}$, we obtain the milder upper bound than low D
because more productions are allowed from the observation.

However, you can easily find that these constraints are obtained only
when $\eta$ is fixed. If we chose the other $\eta$, e.g. which
predicts more D/H than the upper bound of the observation in SBBN,
then, the almost all parameter regions would have been excluded
because both D and $\hefour$ tend to increase in the hadron-injection
scenario. Namely, any constraints, which are obtained when we fix
$\eta$ {\it a priori}, have little meaning. To correctly compare each
model in the various parameters ($\eta$, $\tau_X$, and $n_X/s$), we
should perform the Maximum Likelihood analysis and the $\chi^2$
fitting in wide parameter region including both the observational and
theoretical errors. To estimate the theoretical errors, we perform the
Monte Carlo simulation including the theoretical uncertainties which
come from experimental errors of nuclear reaction and hadron-nucleon
reaction rates.

Concerning the detail of the executions, we have already explained
in~\ref{subsec:SBBN}. For the hadron-nucleon interaction rate, we
adopt 50$\%$ error for each cross section because there are not any
adequate experimental data for the uncertainties of cross
sections. Therefore, we take the larger errors to get a conservative
bound here.

\begin{figure}
      \begin{center}
          \centerline{\psfig{figure=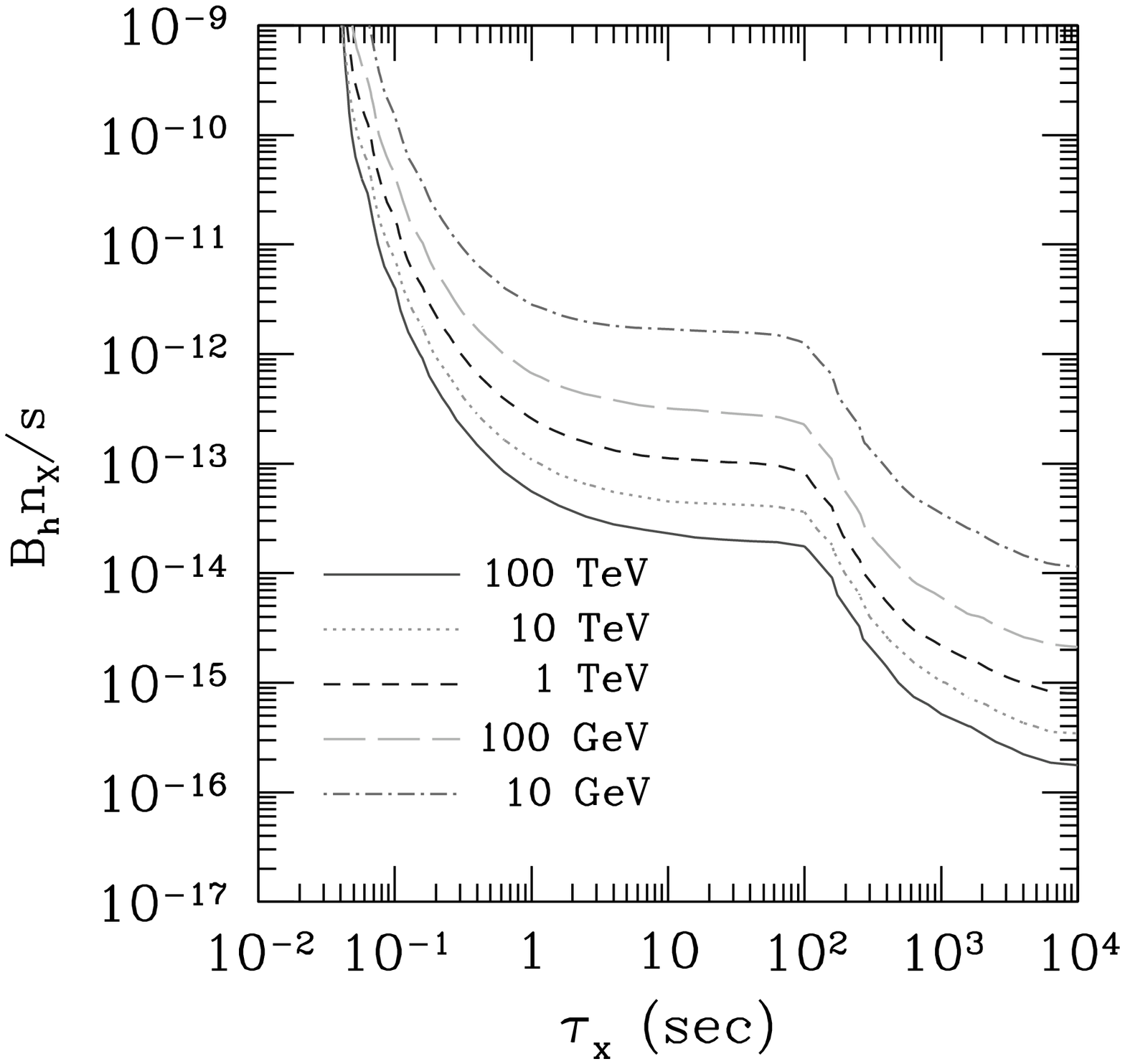,width=17cm}}
          \caption{
          Plot of the contour of the confidence level (C.L.) in
          ($\tau_X, B_hn_X/s$) plane for low D. The region below the
          line is allowed by the observations at 95$\%$ C.L. $\tau_X$
          is the lifetime of $X$, $B_h$ is the branching ratio into
          hadrons, and $n_X/s$ denotes the number density of $X$ per
          entropy density. It is the case that the mass of $X$ is,
          $m_X$ = 100 TeV (solid line), 10 TeV (dotted line), 1 TeV
          (dashed line), 100 GeV (long dashed line), or 10 GeV
          (dot-dashed line) respectively.
          }
          \label{fig:lowd}
      \end{center}
\end{figure}

\begin{figure}
      \begin{center}
          \centerline{\psfig{figure=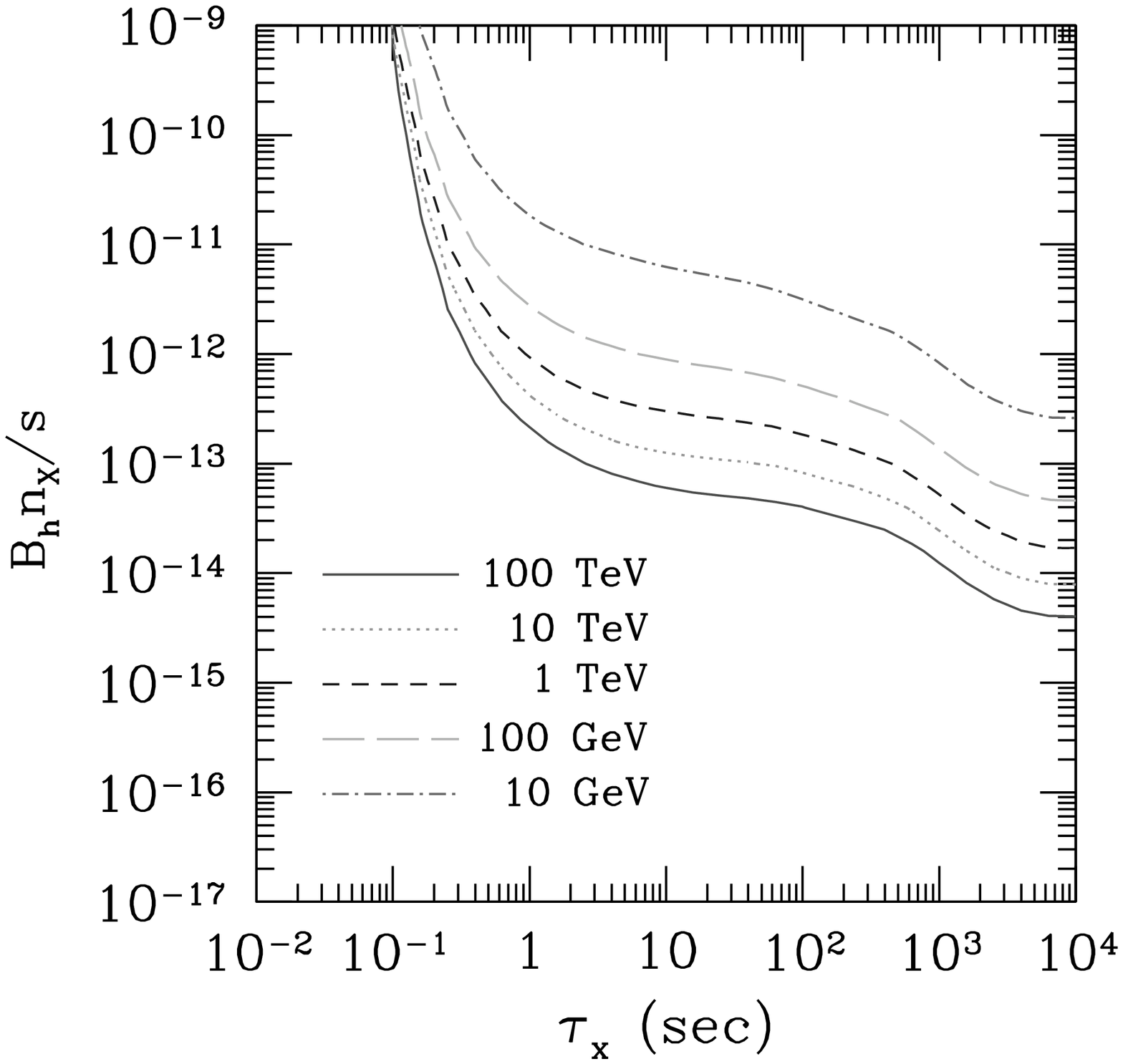,width=17cm}}
          \caption{
          Plot of the contour of the confidence level (C.L.) in
          ($\tau_X, B_hn_X/s$) plane for high D. The region below the
          line is allowed by the observations at 95$\%$ C.L. $\tau_X$
          is the lifetime of $X$, $B_h$ is the branching ratio into
          hadrons, and $n_X/s$ denotes the number density of $X$ per
          entropy density. It is the case of the mass $m_X$ = 100
          TeV (solid line), 10 TeV (dotted line), 1 TeV (dashed line),
          100 GeV (long dashed line), or 10 GeV (dot-dashed line)
          respectively.
          }
          \label{fig:highd}
      \end{center}
\end{figure}

In Fig.~\ref{fig:lowd} we plot the results of the $\chi^2$ fitting at
95$\%$ C.L. in ($\tau_X, B_hn_X/s$) plane projected on $\eta$ axis in
the case of low D which is obtained by performing the Maximum
Likelihood analysis. The region below the line is allowed by the
observations ($\hefour$, D, and $\liseven$) for the various mass of
$X$. If $m_X$ becomes heavier, more hadrons are emitted in the decay,
and the upper bound becomes more stringent. Comparing the case of
$m_X$ = 100 GeV in Fig.~\ref{fig:lowd} with that in
Fig.~\ref{fig:non_MC}, the upper bound obtained in the Monte Carlo
simulation is milder. That is because we did not adopt the naive
2$\sigma$ observational upper bounds with fixed $\eta$, but we
searched the wide range of $\eta$, not forgetting $\tau_X$ and
$n_X/s$, and we performed the Maximum Likelihood analysis to include
both all the observational and theoretical uncertainties. In
Fig.~\ref{fig:highd} we also plot the results of high D. Compared to
the case of low D (Fig.~\ref{fig:lowd}), the obtained upper bound
becomes milder because more D are allowed by the observations in high
D case.

As we also discussed in the previous section, the above treatment
might underestimate the deuterium abundance for $\tau_X \gtrsim 150
\sec$ because deuterium is produced by the destruction of $^4$He by
the high-energy free neutrons in such a relatively late epoch.
Therefore, that means we obtained the conservative limits only for
longer lifetime than 150 sec in this paper.


Here we consider one of the concrete models of the hadronic decay. If
we assume that the parent massive particle is gravitino and that it
mainly decays into a photon and a photino ($\psi_{3/2} \to
\tilde{\gamma} + \gamma$), the lifetime $\tau_{3/2}$ is related to the
gravitino mass $m_{3/2}$ as
\begin{eqnarray}
    \tau_{3/2}\simeq 4\times 10^2{\rm ~sec} \times
    \left(\frac{m_{3/2}}{10 {\rm ~TeV}} \right)^{-3}.
\end{eqnarray}

\begin{figure}
      \begin{center}
          \centerline{\psfig{figure=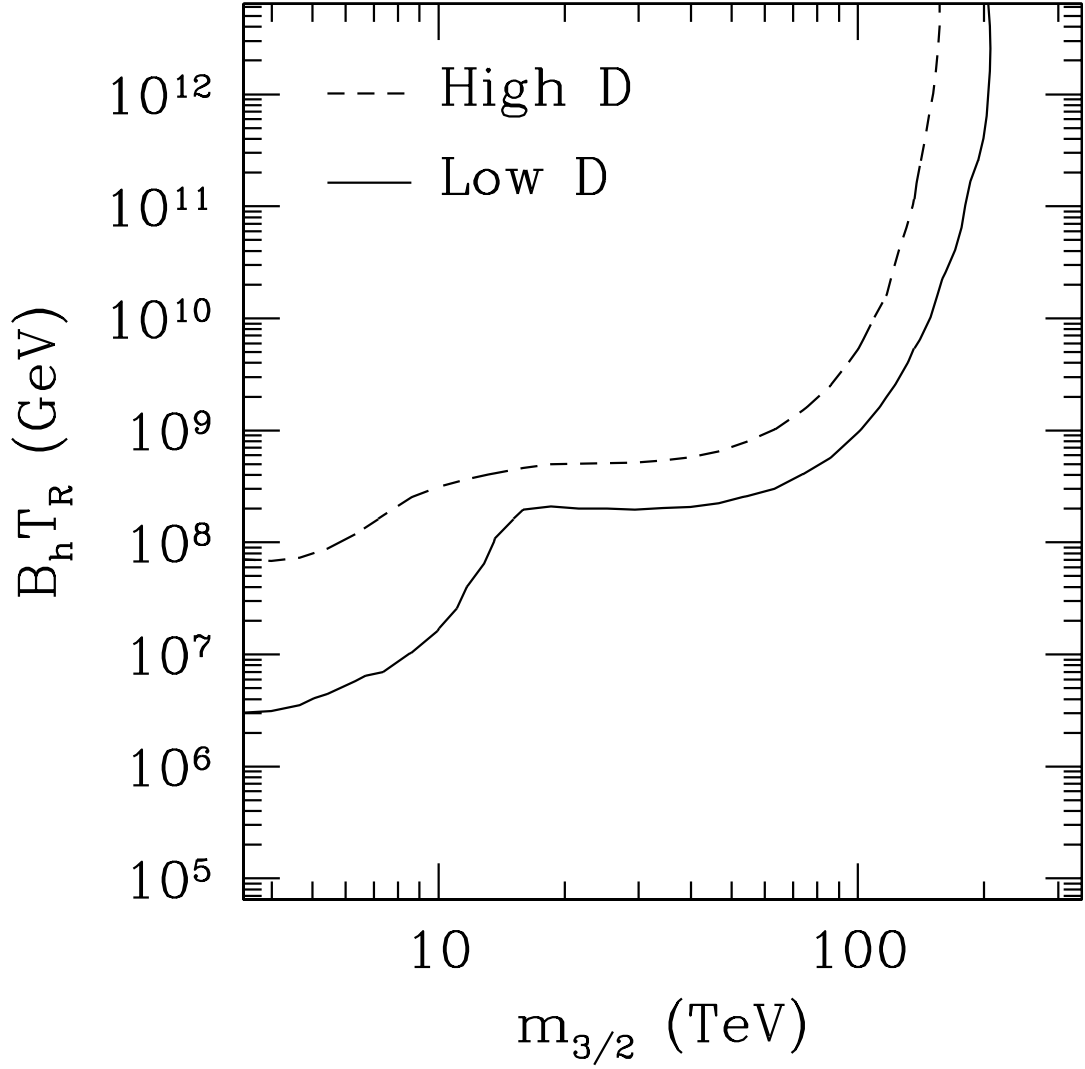,width=17cm}}
          \caption{
          Plot of the upper bound on the reheating temperature after
          inflation at 95$\%$ C.L. as a function of the gravitino mass
          $m_{3/2}$. Here $B_h$ is the branching ratio into hadrons (=
          0.01 -- 1). The solid line (dashed line) denotes the case of
          low D (high D). The region below the line is allowed by the
          observations.
          }
          \label{fig:m_tr}
      \end{center}
\end{figure}

In addition, if we assume that the gravitino is produced through the
thermal scattering in the reheating process after
inflation,\footnote{
In the last two years, it has been claimed that gravitinos might be
also produced in the preheating epoch
non-thermally~\cite{GRT,KKLP,LK}. On the other hand, these days it was
pointed out that such a effect is not important if we realistically
consider two chiral multiplets to distinguish between inflatino and
goldstino~\cite{NPS}, although it may depend on the mixing of the
supersymmetry breaking sector and the inflaton sector~\cite{GRT01}. If
the non-thermal production is effective, however, the obtained
constraint might be severer.
}
we relate the abundance $n_{3/2}/s$ of the gravitino with the
reheating temperature $T_R$~\cite{Kawasaki:1995af},
\begin{eqnarray}
    \frac{n_{3/2}}{s} \simeq 1.6 \times 10^{-12} \times
    \left(\frac{T_R}{10^{10}{\rm GeV}}\right).
\end{eqnarray}
In Fig.~\ref{fig:m_tr} we plot the upper bound on the reheating
temperature after inflation at 95$\%$ C.L. as a function of the
gravitino mass $m_{3/2}$. The solid line (dashed line) denotes the
case of low D (High D). The region below the line is allowed by the
observations. As we discussed before, $B_h$ can become $\sim {\cal
O}$($\alpha$) at least even if the main decay mode is only photons,
because photon has the electromagnetic coupling with $q\anti{q}$, i.e.
($B_h$ = 0.01 -- 1). For $m_{3/2} \lesssim 10 \tev$, they mean the
conservative upper bound.

\section{Summary and Conclusions}
\label{sec:conclusion}

In this paper we have discussed the effects of the late-time hadron
injection on the primordial nucleosynthesis which are caused by the
decay of an unstable massive particle $X$ when the lifetime is
relatively short $10^{-2} \sec \lesssim \tau_X \lesssim 10^4 \sec$.
If the massive particle decays into quarks or gluons, they quickly
fragment into hadrons. Then the high energy hadrons would be emitted
into the electromagnetic thermal bath near the BBN epoch. Because the
background photons and electrons are sufficiently energetic in the
epoch, such high energy hadrons lose their almost all kinetic energies
through the electromagnetic interaction, and they are approximately
stopped before they interact with the background nucleons ($p$ and
$n$) except for neutral kaon $K_L^0$. Then, they scatter off the
background nucleons by the threshold cross sections only for the
exothermic reactions and can extraordinarily inter-convert $p$ and $n$
each other strongly through the hadron-nucleon interaction even after
the freeze-out time of the neutron to proton ratio $n/p$. At that time
it is expected that the background proton tends to be changed into
neutron through the strong interaction since protons are more abundant
than neutrons, and $n/p$ tends to increase. As a result, in
particular, the abundance of $\hefour$ extraordinarily increases
because it is the most sensitive to the freeze out value of
$n/p$. Then, we can constrain the abundance of $X$ and obtain the
informations of $\tau_X$ from the observational light element
abundances.

Here we have studied the hadron injections by using the JETSET 7.4
Monte Carlo event generator~\cite{Sjostrand:1994yb} to quantitatively understand
the hadron jets to agree with the collider
experiments~\cite{PDG}. Thanks to the treatments, we can estimate the
number of the emitted hadrons as a function of the energy of jets,
i.e. as a function of the mass of $X$, even in the regions where
there is no experimental data. In addition we can also obtain the
energy spectrum of the emitted $K^0_L$ for various masses of $X$. This
is very important in the computations because $K^0_L$ is never stopped
in the electromagnetic plasma, and we should know the energy
distributions of $K^0_L$'s.  On the other hand, we also have estimated
the energy-dependent cross sections for $K^0_L$-nucleon scattering
using the existing experimental data~\cite{PDG,durham}. With these
data, we could properly include the hadron-injection effects in BBN
computations.

To estimate the theoretical errors, we performed Monte Carlo
simulation including the theoretical uncertainties which come from
those of the elementary nuclear reaction rates and hadron-nucleon
interaction rates. To obtain the degree of agreements between theory
and observation, we performed the Maximum Likelihood method and the
$\chi^2$ fitting including both the observational and theoretical
errors. To correctly compare each model in the various parameters
($\eta$, $\tau_X$, and $n_X/s$), the above procedure is quite crucial
because a constraint which is obtained when we intentionally fix the
parameters has little meaning.

As a result, we have obtained the upper bound on the abundance $n_X/s$
as a function of the lifetime $\tau_X$ to agree with the observations
for the wide range of the mass $m_X =$ 10 GeV -- 100 TeV which are
relevant for various models of supergravity or superstring theory.
However, we might have underestimated the deuterium abundances where
the lifetime is longer than $\order(10^2) \sec$ because deuterium can
be produced by the destruction of $^4$He by the high-energy free
neutrons, i.e. ``hadro-dissociation'' effects which we ignored in this
work. Therefore, if the process is effective, that would mean we
obtained the conservative upper bounds only for $\tau_X \gtrsim
\order(10^2) \sec$. In the separate paper, we will comprehensively
study the subject~\cite{KKM01}. We have also applied the results
obtained by a generic hadronic decaying particle to gravitino
$\psi_{3/2}$. Then we have got the upper bound on the reheating
temperature after primordial inflation as a function of the mass, $T_R
\lesssim 10^7 - 10^8 \ \gev$ ($T_R \lesssim 10^8 - 10^9 \ \gev$) for
$m_{3/2} = 10 - 100 \ \tev$ at 95 $\%$ C.L. in the case of low D (high
D).

\section{Acknowledgments}

The author wishes to thank T. Asaka, M. Kawasaki, K. Maki, T. Moroi,
and J. Yokoyama for valuable discussions and suggestions.  He also
thanks J. Arafune, O. Biebel, S. Mihara and M.M. Nojiri for useful
comments.



%

\newpage

\newpage

\newpage

\newpage

\newpage

\newpage

\newpage

\newpage

\newpage

\newpage

\newpage

\newpage

\newpage

\newpage

\end{document}